\documentclass[12pt]{article}
\pdfoutput=1

\usepackage[utf8]{inputenc}
\usepackage[left=2.55cm, right=2.55cm, top=2.55cm, bottom=2.55cm]{geometry}
\usepackage{amsmath,amssymb,amsbsy}
\usepackage{slashed}
\usepackage{xcolor}
\usepackage{graphicx}
\usepackage{url}
\usepackage{cancel}
\usepackage{cite}
\usepackage[colorlinks=true,allcolors=darkpurple,pdfborder={0 0 0},linktocpage=false]{hyperref}
\usepackage{tabularx,booktabs}
\usepackage{multicol}
\usepackage{units}
\usepackage{xspace}
\usepackage[labelfont=bf]{caption}
\usepackage[section]{placeins}

\definecolor{darkred}{rgb}{0.6,0,0}
\definecolor{darkpurple}{rgb}{0.5,0,0.5}
\def\vev#1{\left\langle #1\right\rangle}
\def\hc{\text{h.c.}}

\def\neta{n_\eta}
\def\nN{n_N}
\def\Tr{\text{Tr}}

\begin{document}

\vspace*{-2cm}
\begin{flushright}
IFIC/20-13 \\
\vspace*{2mm}
\end{flushright}

\begin{center}
\vspace*{15mm}

\vspace{1cm}
{\Large \bf 
Generalizing the Scotogenic model
} \\
\vspace{1cm}

{\bf Pablo Escribano, Mario Reig, Avelino Vicente}

 \vspace*{.5cm} 
Instituto de F\'{\i}sica Corpuscular (CSIC-Universitat de Val\`{e}ncia), \\
C/ Catedr\'atico Jos\'e Beltr\'an 2, E-46980 Paterna (Valencia), Spain

 \vspace*{.3cm} 
\href{mailto:pablo.escribano@ific.uv.es}{pablo.escribano@ific.uv.es}, \href{mailto:mario.reig@ific.uv.es}{mario.reig@ific.uv.es}, \href{mailto:avelino.vicente@ific.uv.es}{avelino.vicente@ific.uv.es}
\end{center}

\vspace*{10mm}
\begin{abstract}\noindent\normalsize
The Scotogenic model is an economical setup that induces Majorana neutrino masses at the 1-loop level and includes a dark matter candidate. We discuss a generalization of the original Scotogenic model with arbitrary numbers of generations of singlet fermion and inert doublet scalar fields. First, the full form of the light neutrino mass matrix is presented, with some comments on its derivation and with special attention to some particular cases. The behavior of the theory at high energies is explored by solving the Renormalization Group Equations.
\end{abstract}




\section{Introduction}
\label{sec:intro}

The experimental observation of neutrino flavor oscillations
constitutes a milestone in particle physics and proves that the
Standard Model (SM) is an incomplete theory. Although many questions
remain open, such as the Majorana or Dirac nature of neutrinos or the
possible violation of CP in the leptonic sector, the SM must certainly
be extended to include a mechanism that accounts for non-zero neutrino
masses and mixings.

Many neutrino mass models have been proposed along the years. Among
them, radiative models are particularly appealing. After the pioneer
models in the
80's~\cite{Zee:1980ai,Cheng:1980qt,Zee:1985id,Babu:1988ki}, countless
radiative models have been proposed and
studied~\cite{Cai:2017jrq}. The suppression introduced by the loop
factors allows one to accommodate the observed solar and atmospheric
mass scales with sizable couplings and relatively light (TeV scale)
mediators. This typically leads to a richer phenomenology compared to
the usual tree-level scenarios and, in fact, the new mediators may
even be accessible to current colliders. Furthermore, in some
radiative models one can easily address a completely independent
problem: the nature of the dark matter (DM) of the Universe. Discrete
symmetries, connected to the radiative origin of neutrino masses, may
be used to stabilize viable DM candidates, resulting in very
economical scenarios~\cite{Restrepo:2013aga}.

The first and arguably most popular model of this class is the
Scotogenic model~\cite{Ma:2006km}. The addition of just three singlet
fermions and one scalar doublet, as well as a \textit{dark}
$\mathbb{Z}_2$ parity under which these new states are odd, suffices
to simultaneously induce neutrino masses at the 1-loop level and
obtain a weakly-interacting DM candidate.

Since the appearance of the original Scotogenic model, many variations
and extensions have been put forward. These include colored versions
of the
model~\cite{FileviezPerez:2009ud,Liao:2009fm,Reig:2018mdk,Reig:2018ztc}
and versions with additional states and/or symmetries, both in
Dirac~\cite{Farzan:2012sa,Wang:2017mcy,Han:2018zcn,Calle:2018ovc,Ma:2019yfo,Ma:2019iwj,CentellesChulia:2019gic,Jana:2019mez,Jana:2019mgj}
and Majorana
fashion~\cite{Ma:2008cu,Ma:2008ym,Farzan:2009ji,Chen:2009gd,Adulpravitchai:2009re,Farzan:2010mr,Aoki:2011yk,Cai:2011qr,Chen:2011bc,Chao:2012sz,Ma:2012ez,Hirsch:2013ola,Bhattacharya:2013mpa,Ma:2013xqa,Ma:2013nga,Brdar:2013iea,Law:2013saa,Patra:2014sua,Ma:2014eka,Fraser:2014yha,Okada:2015vwh,Chowdhury:2015sla,Diaz:2016udz,Ferreira:2016sbb,Ahriche:2016cio,vonderPahlen:2016cbw,Lu:2016dbc,Merle:2016scw,Rocha-Moran:2016enp,Chowdhury:2016mtl,Fortes:2017ndr,Tang:2017rhv,Guo:2018iix,Rojas:2018wym,Aranda:2018lif,Han:2019lux,Suematsu:2019kst,Kang:2019sab,Pramanick:2019oxb,Nomura:2019lnr,Restrepo:2019ilz,Rojas:2019llr,Avila:2019hhv,Kumar:2019tat,Arbelaez:2020uer}. The
$\mathbb{Z}_2$ parity can also be promoted to a
local~\cite{Ma:2013yga,Yu:2016lof} or global $\rm U(1)$
symmetry~\cite{Kubo:2006rm,Sierra:2014kua,Hagedorn:2018spx,Bonilla:2019ipe},
or to a Peccei-Quinn quasi-symmetry
\cite{Ma:2017zyb,Carvajal:2018ohk,delaVega:2020jcp}. Finally,
Scotogenic-like scenarios have also been combined with, or even
obtained from, extended gauge
symmetries~\cite{Parida:2011wh,Leite:2019grf,Han:2019diw,Wang:2019byi}.

Here we pursue a different type of generalization of the Scotogenic
model. In its original version, three generations of singlet fermions
and a single copy of the inert doublet were included.~\footnote{Even
  though this version of the Scotogenic model is often referred to as
  \textit{the minimal Scototogenic model}, we note that more minimal
  setups can be built~\cite{Farzan:2009ji,Rojas:2018wym,Aranda:2018lif}.} However,
this was just a choice and a Scotogenic model with alternative
numbers of generations can be
considered~\cite{Hehn:2012kz,Fuentes-Martin:2019dxt}. This is the aim
of this paper, to introduce the \textit{general Scotogenic model},
with arbitrary numbers of generations of the Scotogenic states, and
study its more relevant features.

The rest of the manuscript is organized as follows. In
Sec.~\ref{sec:scot} we present our generalization of the Scotogenic
model to any number of singlet fermions and inert scalar
doublets. Sec.~\ref{sec:numass} is devoted to the calculation of the
induced 1-loop neutrino masses, whereas some aspects of the
high-energy behavior of the model and the relevance of thermal effects
are discussed in Secs.~\ref{sec:running} and \ref{sec:thermal},
respectively. We summarize our findings and conclude with some further
comments in Sec.~\ref{sec:conclusions}. Additional details are given
in Appendices~\ref{sec:app1} and \ref{sec:app2}.

\section{The general Scotogenic model}
\label{sec:scot}

The Scotogenic model~\cite{Ma:2006km} is a simple extension of the SM
that induces radiative neutrino masses and provides a potential dark
matter candidate. Here we consider a generalization of the model. The
SM particle content is extended by an unspecified number, $\nN$, of
singlet fermions $N$, and also an arbitrary number, $\neta$, of inert
scalar doublets $\eta$. Particular cases of this particle spectrum can
be labeled by their $(\nN,\neta)$ values. In addition, the symmetry
group of the SM is enlarged with a \emph{dark} $\mathbb{Z}_2$ parity,
under which all the new fields are odd, while the SM particles are
even. The scalar and fermion particle content of the model, as well as
their representations under the gauge group $\rm SU(3)_c \times
SU(2)_L \times U(1)_Y$ and the $\mathbb{Z}_2$ parity of the model are
given in Tab.~\ref{tab:bmlnur}.

{
\renewcommand{\arraystretch}{1.6}
\begin{table}[!h]
	\centering
	\begin{tabular}{|c|c||ccc|c|}
		\hline
		Field & Generations & $\rm SU(3)_c$ & $\rm SU(2)_L$ & $\rm U(1)_Y$ & $\mathbb{Z}_2$\\
                \hline
		$\ell_L$ & $3$ & $\mathbf{1}$ & $\mathbf{2}$ & $-1/2$ & $+$ \\
                $e_R$ & $3$ & $\mathbf{1}$ & $\mathbf{1}$ & $-1$ & $+$ \\
                $H$ & $1$ & $\mathbf{1}$ & $\mathbf{2}$ & $1/2$ & $+$ \\ 
                \hline		
		$\eta$ & $\neta$ & $\mathbf{1}$ & $\mathbf{2}$ & $1/2$ & $-$ \\
                $N$ & $\nN$ & $\mathbf{1}$ & $\mathbf{1}$ & $0$ & $-$ \\
		\hline
	\end{tabular}
	\caption{Scalar and fermion particle content of the model and
          representations under the gauge and global
          symmetries. $\ell_L$ and $e_R$ are the SM left- and
          right-handed leptons, respectively, and $H$ is the SM Higgs
          doublet.
	\label{tab:bmlnur}}
\end{table}
}

The relevant Yukawa and bare mass terms for our discussion are
\begin{equation} \label{eq:yukawa}
\mathcal{L}_N \supset y_{n a \alpha} \, \overline{N}_n \, \eta_a \, \ell_L^\alpha + \frac{1}{2} \, M_{N_n} \, \overline{N^c}_n \, N_n + \hc \, ,
\end{equation}
where $n=1,\dots,\nN$, $a=1,\dots,\neta$ and $\alpha=1,2,3$ are
generation indices and $y$ is a general complex $\nN \times \neta
\times 3$ object. Besides, $M_N$ is a symmetric $\nN \times \nN$
Majorana mass matrix that has been chosen diagonal without loss of
generality. Furthermore, one can also write the scalar potential
\begin{equation} 
  \begin{split}
    \mathcal{V} &=m_{H}^{2} H^{\dagger} H+\left(m_{\eta}^{2}\right)_{a b} \eta_{a}^{\dagger} \eta_{b}+\frac{1}{2} \, \lambda_{1}\left(H^{\dagger} H\right)^{2}+\frac{1}{2} \, \lambda_{2}^{a b c d}\left(\eta_{a}^{\dagger} \eta_{b}\right)\left(\eta_{c}^{\dagger} \eta_{d}\right) \\ 
    &+\lambda_{3}^{a b}\left(H^{\dagger} H\right)\left(\eta_{a}^{\dagger} \eta_{b}\right)+\lambda_{4}^{a b}\left(H^{\dagger} \eta_{a}\right)\left(\eta_{b}^{\dagger} H\right) \\ 
    &+\frac{1}{2} \left[\lambda_{5}^{a b} \left(H^{\dagger} \eta_{a}\right)\left(H^{\dagger} \eta_{b}\right) + \, \hc \right] \, .
  \end{split}
\label{eq:potential}
\end{equation}
Here all the indices are $\eta$ generation indices. Therefore,
$m_\eta^2$ and $\lambda_{3,4,5}$ are $\neta \times \neta$ matrices
while $\lambda_2$ is an $\neta \times \neta \times \neta \times
\neta$ object. Note that $\lambda_5$ must be symmetric whereas
$\lambda_{3,4}$ must be Hermitian. Again, $m_\eta^2$ will be assumed
to be diagonal without loss of generality. Finally, we highlight the
presence of the scalar potential quartic couplings $\lambda_5^{a b}$,
which play a major role in the neutrino mass generation mechanism, as
shown in Sec.~\ref{sec:numass}.

We will assume that the minimization of the scalar potential in
Eq. \eqref{eq:potential} leads to the vacuum configuration
\begin{equation} \label{eq:VEVs}
\vev{H^0} = \frac{v}{\sqrt{2}} \quad , \quad \vev{\eta_a^0} = 0 \, ,
\end{equation}
with $a = 1, \dots, \neta$. Therefore, only the neutral component of
$H$ acquires a non-zero vacuum expectation value (VEV), which breaks
the electroweak symmetry in the standard way, while the $\eta_a$
scalars are inert doublets with vanishing VEVs. In this way, the
$\mathbb{Z}_2$ symmetry remains unbroken and the stability of the
lightest $\mathbb{Z}_2$-charged particle is guaranteed. We will come
back to the possibility of $\mathbb{Z}_2$ breaking due to
Renormalization Group Equations (RGEs) effects later.

We now decompose the
neutral component of the $\eta_a$ multiplets, $\eta_a^0$, as
\begin{equation}
	\eta_{a}^{0}=\frac{1}{\sqrt{2}} \, \left(\eta_{R_{a}} + i \, \eta_{I_{a}} \right) \, .
\end{equation}
In the following we will assume that all the parameters in the scalar
potential are real, hence conserving CP in the scalar sector. In this
case, the real and imaginary components of $\eta_a^0$ do not
mix. After electroweak symmetry breaking, the $\neta \times \neta$
mass matrices for the real and imaginary components are given by
\begin{equation}
(\mathcal{M}_{R}^{2})_{a b} = (m_{\eta})^{2}_{a a} \, \delta_{a b} + \left( \lambda_{3}^{a b} + \lambda_{4}^{a b} + \lambda_5^{a b} \right) \, \frac{v^2}{2}
\end{equation}
and
\begin{equation}
(\mathcal{M}_{I}^{2})_{a b} = (m_{\eta})^{2}_{a a} \, \delta_{a b} + \left( \lambda_{3}^{a b} + \lambda_{4}^{a b} - \lambda_5^{a b} \right) \, \frac{v^2}{2} \, ,
\end{equation}
respectively. We note that $\mathcal{M}_R^2 = \mathcal{M}_I^2$ in the
limit $\lambda_5 \to 0$, in which all the elements of $\lambda_5$
vanish. This will be crucial in the calculation of neutrino masses, as
shown below. Both mass matrices can be brought into diagonal form by
means of a change of basis. The gauge eigenstates, $\eta_{A_a}$, are
related to the mass eigenstates, $\hat{\eta}_{A_b}$, where $A = R, I$,
by
\begin{equation}
  \eta_A = V_A \, \hat{\eta}_A \, .
\end{equation}
Here $\eta_{A}$ and $\hat{\eta}_{A}$ are $\neta$-component vectors. In
general, the $\neta \times \neta$ matrices $V_{A}$ are unitary, such
that $V_A V_A^\dagger = V_A^\dagger V_A = \mathbb{I}_{\neta}$, where
$\mathbb{I}_{\neta}$ is the $n_\eta \times n_\eta$ identity
matrix. However, in the simplified scenario of CP conservation in the
scalar sector, $\mathcal{M}_{R}^{2}$ and $\mathcal{M}_{I}^{2}$ are
real symmetric matrices, and then the $V_{A}$ matrices are orthogonal,
such that $V_A V_A^T = V_A^T V_A = \mathbb{I}_{\neta}$. With these
transformations, the diagonal mass matrices are given by
\begin{equation}
\widehat{\mathcal{M}}_{A}^{2} = \left( \begin{array} {ccc} 
  {m_{A_1}^{2}} & & {0} \\ 
  & \ddots & \\
  {0} & & {m_{A_{\neta}}^{2}} \end{array} \right) = V_{A}^{T} \mathcal{M}_{A}^{2} V_{A} \, .
\end{equation}
The resulting analytical expressions for the mass eigenvalues
$m_{A_{a}}^{2}$ and mixing matrices $V_A$ involve complicated
combinations of the scalar potencial parameters. However, under the
assumptions\footnote{Note that this assumption is technically
  natural~\cite{tHooft:1979rat}: the smallness of $\lambda_5$ is not
  dynamically explained but is stable against RGE flow. This is due to
  the fact that the limit $\lambda_5 \to 0$ increases the symmetry of
  the model by restoring lepton number. Therefore, if $\lambda_5$ is
  set small at one scale it will remain small at all scales.}
\begin{equation} \label{eq:assumptions}
\lambda_{3,4}^{a a} \, \frac{v^2}{2} \ll \left(m_{\eta}^{2}\right)_{a a} \quad \text{and} \quad \lambda_{5}^{a b} \ll \lambda_{3,4}^{a b} \ll 1
\end{equation}
one can find simple expressions. The $m_{A_{a}}^{2}$ mass eigenvalues
are given by
\begin{align}
  \label{eq:mass_assumptions1}
  m_{R_{a}}^{2} &= \left(m_{\eta}^{2}\right)_{a a} + \left( \lambda_{3}^{a a} + \lambda_{4}^{a a} + \lambda_{5}^{a a} \right) \, \frac{v^2}{2} \, , \\
  m_{I_{a}}^{2} &= \left(m_{\eta}^{2}\right)_{a a} + \left( \lambda_{3}^{a a} + \lambda_{4}^{a a} - \lambda_{5}^{a a} \right) \, \frac{v^2}{2} \, .
  \label{eq:mass_assumptions2}
\end{align}
We note that the mass splitting $m_{R_a}^2 - m_{I_a}^2 =
\lambda_5^{aa} \, v^2$ vanishes in the limit $\lambda_5 \to 0$. In
what concerns the $V_A$ orthogonal matrices, each of them can be
expressed as a product of $n_\eta (n_\eta - 1)/2$ rotation matrices,
with the scalar mixing angles given by
\begin{equation} \label{eq:angles}
\tan 2 \, \theta_A^{ab} = \frac{2 \, (\mathcal{M}_{A}^{2})_{a b}}{(\mathcal{M}_{A}^{2})_{b b} - (\mathcal{M}_{A}^{2})_{a a}} = \left( \lambda_3^{ab} + \lambda_4^{ab} + \kappa_A^2 \, \lambda_5^{ab} \right) \, \frac{v^2}{m_{A_{b}}^{2} - m_{A_{a}}^{2}} \, ,
\end{equation}
where the $\kappa_A^2$ sign ($\kappa_R^2 = +1$ and $\kappa_I^2 = -1$)
has been introduced.

\section{Neutrino masses}
\label{sec:numass}

\begin{figure}[h!]
\centering
\includegraphics[width=0.48\linewidth,keepaspectratio]{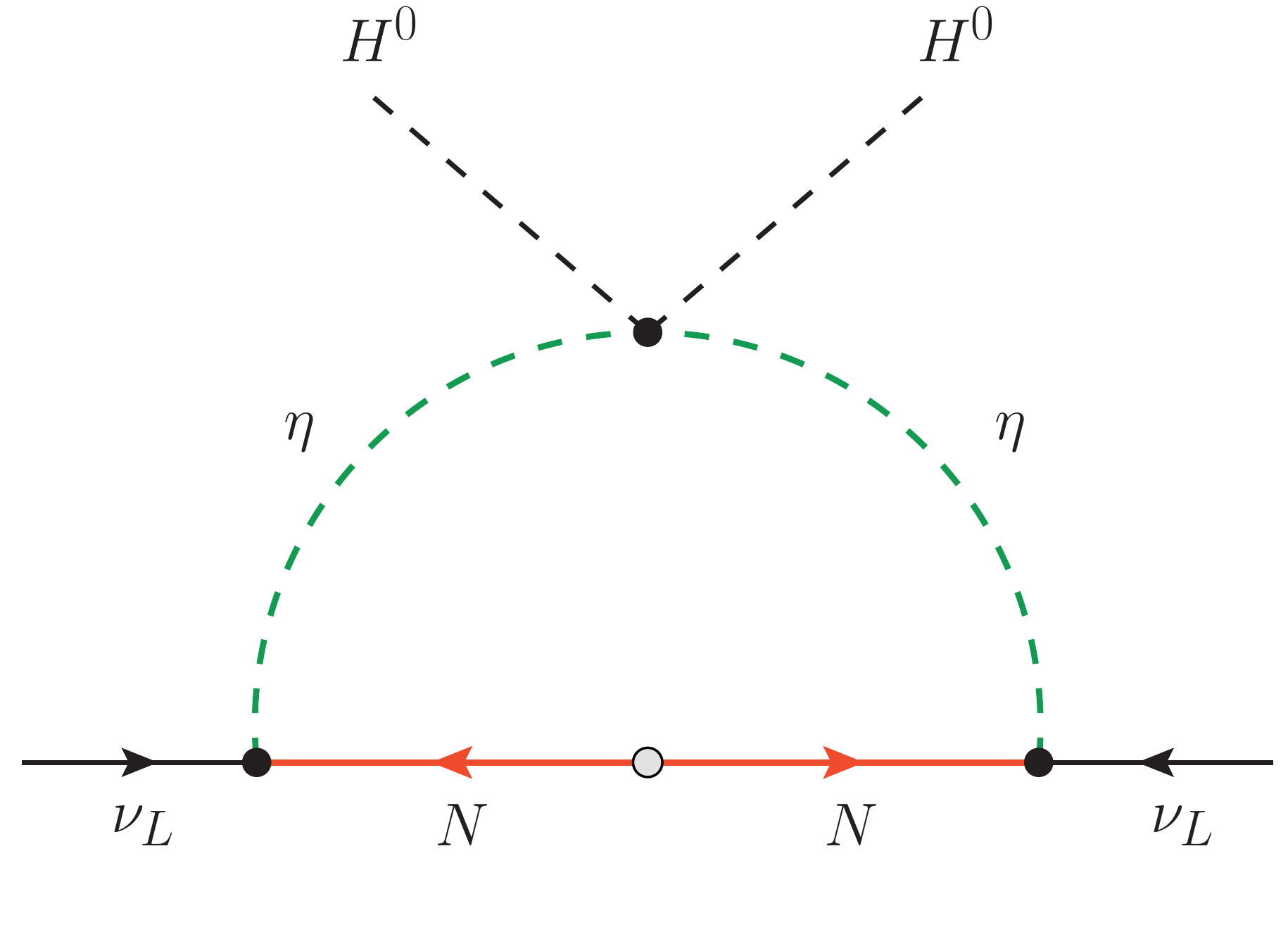}
\includegraphics[width=0.48\linewidth,keepaspectratio,trim= 0in -0.13in 0in 0in]{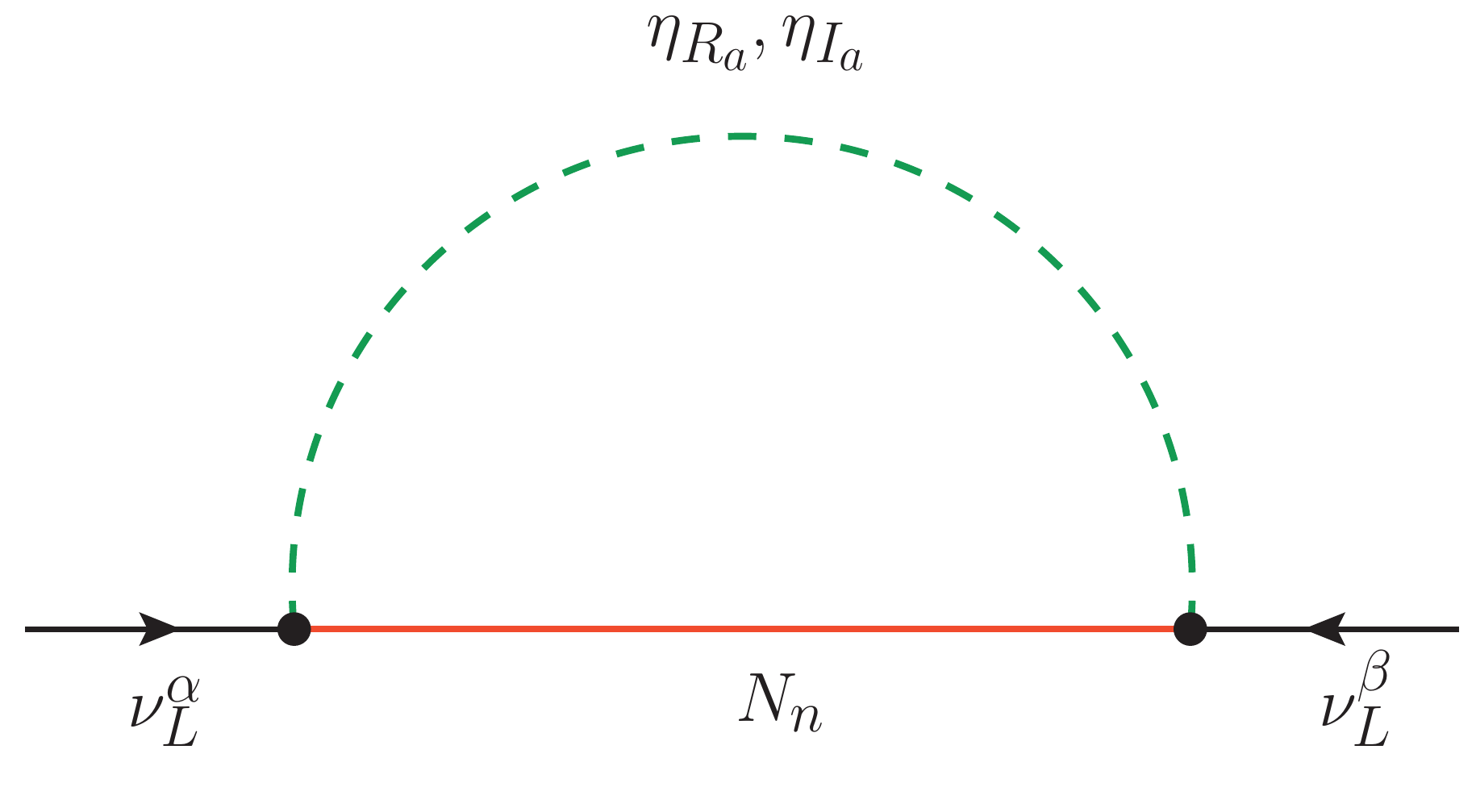}
\caption{Neutrino mass generation. To the left, Feynman diagram with gauge eigenstates. To the right, the analogous Feynman diagram with the physical mass eigenstates that propagate in the loop.
\label{fig:numass}}
\end{figure}

The generation of neutrino masses takes place at the 1-loop level
\textit{\`a la scotogenic}~\cite{Ma:2006km}. In the presence of the
terms given in Eqs.~\eqref{eq:yukawa} and \eqref{eq:potential}, lepton
number is explicitly broken in two units, hence inducing Majorana
neutrino masses. Assuming that the potential is such that the $\eta_a$
scalars do not get VEVs, see Eq.~\eqref{eq:VEVs}, neutrino masses are
forbidden at tree-level.  Nevertheless, they are induced at the 1-loop
level, as shown in Fig.~\ref{fig:numass}. Several diagrams contribute
to the neutrino mass matrix. Therefore, one can write
\begin{equation} \label{eq:mnu1}
\left(m_\nu\right)_{\alpha \beta} = \sum_{A,a,n} \left(m_\nu^A\right)_{\alpha \beta}^{a n} \, ,
\end{equation}
where $\left(m_\nu^A\right)_{\alpha \beta}^{a n}$ is the contribution
to $\left(m_\nu\right)_{\alpha \beta}$ generated by the $N_n -
\eta_{A_a}$ loop, given by
\begin{equation} \label{eq:integral}
- i \left( m_\nu^A \right)_{\alpha \beta}^{a n} = C^A_{n a \alpha} \, \int \frac{\text{d}^D k}{\left( 2 \pi \right)^D} \frac{i}{k^2 - m^2_{A_a}} \frac{i \left( \slashed{k} + M_{N_n} \right)}{k^2 - M^2_{N_n}} \ C^A_{n a \beta} \, ,
\end{equation}
where $D = 4 - \varepsilon$ is the number of space-time dimensions, the
external neutrinos are taken at rest and $k$ is the momentum running
in the loop. We note that the term proportional to $\slashed{k}$ does
not contribute because it is odd in the loop momentum. $C^A_{n a
  \alpha}$ is the $N_n - \eta_{A_a} - \nu_L^\alpha$ coupling, given by
\begin{equation} \label{eq:Netanucoup}
C^A_{n a \alpha} = i \, \frac{\kappa_A}{\sqrt{2}} \, \sum_b \, \left( V_A \right)^*_{b a} \, y_{n b \alpha} \, ,
\end{equation}
with $\kappa_R = 1$ and $\kappa_I = i$. Since we assume real
parameters in the scalar sector, complex conjugation in $V_A$ will be
dropped in the following. Replacing Eq.~\eqref{eq:Netanucoup} into
Eq.~\eqref{eq:integral} and introducing the standard Passarino-Veltman
loop function $B_0$~\cite{Passarino:1978jh},

\begin{equation}
B_0 \left( 0, m_{A_a}^2, M_{N_n}^2  \right) = \Delta_\varepsilon + 1 - \frac{m_{A_a}^2 \log m_{A_a}^2 - M_{N_n}^2 \log M_{N_n}^2}{m_{A_a}^2 - M_{N_n}^2},
\end{equation}
where $\Delta_\varepsilon$ diverges in the limit $\varepsilon \to 0$,
Eq.~\eqref{eq:mnu1} becomes
\begin{equation} \label{eq:mnu2}
\boxed{
\left( m_\nu \right)_{\alpha \beta} = - \frac{1}{32 \pi^2} \sum_{A, a, b, c, n} \, M_{N_n} \, \kappa_A^2 \left( V_A \right)_{b a} \, \left( V_A \right)_{c a} \, y_{n b \alpha} \, y_{n c \beta} \, B_0(0, m_{A_a}^2, M_{N_n}) \, . 
}
\end{equation}
Eq.~\eqref{eq:mnu2} constitutes our central result for the 1-loop
neutrino mass matrix in the model. It is important to note that the
divergent pieces cancel exactly. Indeed, the $\kappa_A^2$ factor
implies that the term proportional to $\Delta_\varepsilon$ in
Eq.~\eqref{eq:mnu2} involves the combination
\begin{equation}
\sum_a \left[ \left(V_R\right)_{b a} \ \left(V_R\right)_{c a} - \left(V_I\right)_{b a} \ \left(V_I\right)_{c a} \right] = \left(V_R \, V_R^T \right)_{b c} - \left(V_I \, V_I^T \right)_{b c} = \delta_{bc} - \delta_{bc} = 0 \, ,
\end{equation}
which vanishes due to the orthogonality of the $V_A$ matrices,
ensuring the cancellation of the divergent part of the $B_0$
functions. This was expected since the neutrino mass matrix is
physical and therefore finite.

While Eq.~\eqref{eq:mnu2} provides a simple analytical expression for
the neutrino mass matrix, the dependence on the fundamental parameters
of the model is not explicit. The neutrino mass matrix involves a
product of $V_A$ matrices and $B_0$ functions, both in general
depending on the scalar potential parameters in a non-trivial way. In
order to identify more clearly the role of the scalar potential
parameters, we will work under the assumptions in
Eq.~\eqref{eq:assumptions} and derive an approximate form for the
neutrino mass matrix, valid for small $\lambda_5^{a b}$ couplings and
small mixing angles in the scalar sector. First, it is convenient to
make an expansion in powers of $\lambda_5^{a b} \ll 1$. One can write
\begin{align}
  \left( m_\nu \right)_{\alpha \beta} = - \frac{1}{32 \pi^2} & \sum_{n} \ M_{N_n} \ \sum_{a, b, c} \ y_{n b \alpha} \ y_{n c \beta} 
\label{eq:mnufirstorder} \\  
  & \left\{ \left[ \left( V \right)_{b a} \ \left( V \right)_{c a} \right]^{(0)} \ \left[ B_0^{(1)}(0, m_{R_a}^2, M_{N_n}) - B_0^{(1)}(0, m_{I_a}^2, M_{N_n}) \right] \right. \nonumber \\
  & + \left. \left[ \left( V_R \right)_{b a} \ \left( V_R \right)_{c a} - \left( V_I \right)_{b a} \ \left( V_I \right)_{c a} \right]^{(1)} \ B_0^{(0)}(0, m_{a}^2, M_{N_n}) \right\} + \mathcal{O} \left( \lambda_5^2 \right) \, , \nonumber
\end{align}
where the superindex $^{(i)}$, with $i = 0,1$, denotes the order in
$\lambda_5^{a b}$. We highlight that the expansion begins at 1st order
in $\lambda_5$. This was indeed expected, since $\lambda_5 = 0$ would
imply the restoration of lepton number and massless neutrinos. With
this in mind, the origin of the two terms in
Eq.~\eqref{eq:mnufirstorder} is easy to understand. In the first term,
the $\lambda_5^{ab}$ couplings are neglected in the $V_A$ matrices but
kept at leading order in the $B_0$ functions. This term is
proportional to the $B_0(0, m_{R_a}^2, M_{N_n}) - B_0(0, m_{I_a}^2,
M_{N_n})$ difference, which would vanish for $\lambda_5^{aa} = 0$, see
Eqs.~\eqref{eq:mass_assumptions1} and
\eqref{eq:mass_assumptions2}. The mass matrices for the real and
imaginary components of $\eta^0$ are equal at 0th order in
$\lambda_5$, $\widehat{\mathcal{M}}_{R}^{2 \, (0)} =
\widehat{\mathcal{M}}_{I}^{2 \, (0)}$, and then we can define $V
\equiv V_R^{(0)} = V_I^{(0)}$. In the second term, the
$\lambda_5^{ab}$ couplings are neglected in the $B_0$ functions but
kept at leading order in the $V_A$ mixing matrices. Since
$m_{R_a}^{(0)} =m_{I_a}^{(0)} \equiv m_{a}$ at 0th order in
$\lambda_5^{aa}$, then the $B_0^{(0)}$ function has the argument
\begin{equation}
  m_{a}^{2} = \left(m_{\eta}^{2}\right)_{a a} + \left( \lambda_{3}^{a
    a} + \lambda_{4}^{a a} \right) \, \frac{v^2}{2} \, .
\end{equation}
We note that this term will only be non-zero when the $\lambda_5$
matrix contains non-vanishing off-diagonal entries, since this is the
only way the $\left( V_R \right)_{b a} \ \left( V_R \right)_{c a} -
\left( V_I \right)_{b a} \ \left( V_I \right)_{c a}$ would not vanish
at 1st order in $\lambda_5$. Next, we find approximate expressions for
the $V_A$ mixing matrices. This is only feasible by assuming small
scalar mixing angles, in agreement with Eq.~\eqref{eq:assumptions}. In
this case one can expand $V$ not only in powers of $\lambda_5$, but
also in powers of the small parameter
\begin{equation}
s_{ab} = \frac{1}{2} \, \left( \lambda_3^{ab} + \lambda_4^{ab} \right) \, \frac{v^2}{m_{b}^{2} - m_{a}^{2}} \ll 1 \, ,
\end{equation}
which is defined for $a \ne b$ and corresponds to $\sin \theta_R^{ab}$
or $\sin \theta_I^{ab}$ at 0th order in $\lambda_5$, see
Eq.~\eqref{eq:angles}. With this definition, one finds the general
expression $\left( V \right)_{a b} = \delta_{ab} + (1-\delta_{ab}) \,
s_{ab} + \mathcal{O} \left( s^2 \right)$. Analogous expressions are
found for $V_R$ and $V_I$ replacing $s$ by $\sin \theta_R$ and $\sin
\theta_I$, respectively.  With all these ingredients,
Eq.~\eqref{eq:mnufirstorder} can be written as
\begin{equation} \label{eq:mnu3}
\boxed{
  \left( m_\nu \right)_{\alpha \beta} = \frac{v^2}{32 \pi^2} \sum_{n, a, b} \frac{y_{n a \alpha} \, y_{n b \beta}}{M_{N_n}} \, \Gamma_{abn} + \mathcal{O} \left( \lambda_5^2 \right) + \mathcal{O} \left( \lambda_5 \, s^2 \right) \, ,
}
\end{equation}
where we have defined the dimensionless quantity
\begin{equation} \label{eq:Gamma}
\Gamma_{abn} = \delta_{ab} \, \lambda_5^{aa} \, f_{an} - (1 - \delta_{ab}) \left[ \left( \lambda_5^{aa} \, f_{an} - \lambda_5^{bb} \, f_{bn} \right) \, s_{ab} - \frac{M^2_{N_n}}{m_b^2 - m_a^2} \, \lambda_5^{ab} \ g_{abn} \right]
\end{equation}
and the loop functions
\begin{align}
  f_{an} &= \frac{M_{N_n}^{2}}{m_a^2 - M_{N_n}^{2}} + \frac{M_{N_n}^{4}}{\left(m_a^2 - M_{N_n}^{2}\right)^{2}} \log \frac{M_{N_n}^{2}}{m_a^2} \, , \\
  g_{abn} &= \frac{m_a^2}{m_a^2 - M_{N_n}^2} \log \frac{M_{N_n}^2}{m_a^2} - \frac{m_b^2}{m_b^2 - M_{N_n}^2} \log \frac{M_{N_n}^2}{m_b^2} \, .
\end{align}
Eq.~\eqref{eq:mnu3} involves the quantity $\Gamma_{abn}$, which we
have written in Eq.~\eqref{eq:Gamma} as the sum of two terms. The
first term in $\Gamma_{abn}$ contributes only for $a=b$ and involves
only diagonal elements of $\lambda_5$. The second term, which involves
diagonal as well as off-diagonal elements of $\lambda_5$, only
contributes for $a \ne b$. We also note that $g_{abn} = - g_{ban}$.

Eq.~\eqref{eq:mnu3} is the main analytical result of our work. Under
the assumptions of Eq.~\eqref{eq:assumptions}, it reproduces the
neutrino mass matrix in very good approximation. It is valid for any
$n_N$ and $n_\eta$ values. We will now show how in some particular
cases it reduces to well-known expressions in the literature.

\subsection{Particular case 1: $\boldsymbol{(n_N,n_\eta) = (3,1)}$}

The first example we consider is the standard Scotogenic model
originally introduced in \cite{Ma:2006km} and obtained for
$(n_N,n_\eta) = (3,1)$. In this case, only one inert doublet $\eta$ is
introduced. Therefore all the matrices in the scalar sector become
just scalar parameters: $V_A = 1$, $\lambda_5^{ab} \equiv
\lambda_5^{11} \equiv \lambda_5$ and $(m_\eta^2)_{aa} \equiv
(m_\eta^2)_{11} \equiv m_\eta^2$. Besides, the Yukawa couplings become
$3 \times 3$ matrices: $y_{n a \alpha} \equiv y_{n 1 \alpha} \equiv
y_{n \alpha}$. Similarly, $f_{a n} \equiv f_{1 n} \equiv f_n$, and the
second term in Eq.~\eqref{eq:Gamma} does not contribute. With these
simplifications, the general $\Gamma_{abn}$ reduces to
$\Gamma_{n}^{(3,1)}$, given by
\begin{equation} \label{eq:Gamma31}
\Gamma_{abn}^{(3,1)} \equiv \Gamma_{11n}^{(3,1)} \equiv \Gamma_{n}^{(3,1)} = \lambda_5 \, f_{n} \, .
\end{equation}
Replacing this into Eq.~\eqref{eq:mnu3}, one obtains the well-known
neutrino mass matrix
\begin{equation} \label{eq:mnu31}
\left(m_{\nu}\right)_{\alpha \beta}^{(3,1)} = \frac{\lambda_{5} \, v^{2}}{32 \pi^{2}} \sum_{n} \frac{y_{n \alpha} \, y_{n \beta}}{M_{N_n}} \left[ \frac{M_{N_n}^{2}}{m_{0}^{2} - M_{N_n}^{2}} + \frac{M_{N_n}^{4}}{\left(m_{0}^{2} - M_{N_n}^{2}\right)^{2}} \log \frac{M_{N_n}^{2}}{m_{0}^{2}} \right] \, ,
\end{equation}
with $m_0^2 = m_\eta^2 + (\lambda_3 + \lambda_4) \, v^2 / 2$. This
expression agrees with \cite{Ma:2006km} up to a factor of $1/2$ that
was missing in the original reference.~\footnote{The correct
  expression was first shown in version 1 of \cite{Merle:2015gea} and
  later reproduced in \cite{Vicente:2015zba,Cai:2017jrq}.}

\subsection{Particular case 2: $\boldsymbol{(n_N,n_\eta) = (1,2)}$}

A version of the Scotogenic model with one singlet fermion and two
inert doublets, $(n_N,n_\eta) = (1,2)$, has been considered in
\cite{Hehn:2012kz,Fuentes-Martin:2019dxt}. Since the model contains
only one singlet fermion $N$, $M_{N_n} \equiv M_N$ is just a
parameter. The Yukawa couplings become $2 \times 3$ matrices: $y_{n a
  \alpha} \equiv y_{1 a \alpha} \equiv y_{a \alpha}$. Finally, $f_{n
  a} \equiv f_{1 a} \equiv f_a$ and $g_{abn} \equiv g_{ab1} \equiv
g_{ab}$. Both references work in the basis in which the $m_\eta^2$
matrix is diagonal. However, they take different simplifying
assumptions about the scalar potential parameters.

In~\cite{Hehn:2012kz} the matrix $\lambda_3 + \lambda_4$ was assumed
to be diagonal. In this case, which we denote as scenario $(1,2)_{\rm
  \, I}$, $(1 - \delta_{ab}) s_{ab} = 0$ and the general
$\Gamma_{abn}$ reduces to
\begin{equation} \label{eq:Gamma12a}
\Gamma_{abn}^{(1,2)_{\rm \, I}} \equiv \Gamma_{ab1}^{(1,2)_{\rm \, I}} \equiv \Gamma_{ab}^{(1,2)_{\rm \, I}} = \delta_{ab} \, \lambda_5^{aa} \, f_{an} + (1 - \delta_{ab}) \, \frac{M^2_{N_n}}{m_b^2 - m_a^2} \, \lambda_5^{ab} \ g_{abn} \, .
\end{equation}
Replacing this expression into Eq.~\eqref{eq:mnu3} and arranging the
different pieces properly, one obtains
\begin{equation} \label{eq:mnu12I}
\left( m_\nu \right)_{\alpha \beta}^{(1,2)_{\rm \, I}} = \frac{v^2}{32 \pi^2} \sum_{a,b} y_{a \alpha} \, y_{b \beta} \, \lambda_{5}^{a b} \, \frac{M_N}{m_{b}^{2} - M_N^{2}} \left[ \frac{m_{b}^{2}}{m_{a}^{2}-m_{b}^{2}} \log \frac{m_{a}^{2}}{m_{b}^{2}}-\frac{M_N^{2}}{m_{a}^{2}-M_N^{2}} \log \frac{m_{a}^{2}}{M_N^{2}} \right] \, ,
\end{equation}
which agrees with the result in \cite{Hehn:2012kz} up to a global
factor of $1/4$.

On the other hand, a diagonal $\lambda_5$ matrix was taken
in~\cite{Fuentes-Martin:2019dxt}. We denote this as scenario
$(1,2)_{\rm \, II}$. Again, this simplifies $\Gamma_{abn}$, which
becomes
\begin{equation} \label{eq:Gamma12b}
\Gamma_{abn}^{(1,2)_{\rm \, II}} \equiv \Gamma_{ab1}^{(1,2)_{\rm \, II}} \equiv \Gamma_{ab}^{(1,2)_{\rm \, II}} = \delta_{ab} \, \lambda_5^{aa} \, f_{an} - (1 - \delta_{ab}) \left( \lambda_5^{aa} \, f_{an} - \lambda_5^{bb} \, f_{bn} \right) \, s_{ab} \, .
\end{equation}
With this result, one can easily use Eq.~\eqref{eq:mnu3} to derive
\begin{equation} \label{eq:mnu12II}
\left( m_{\nu} \right)_{ \alpha \beta}^{(1,2)_{\rm \, II}} = \frac{v^{2}}{32 \pi^{2} M_{N}} \, \sum_{a,b,c} \, y_{a \alpha} \, y_{b \beta} \, \lambda_5^{cc} \, f_c \, X_{abc} \, , 
\end{equation}
with
\begin{equation}
X_{abc} = \delta_{ab} \delta_{bc} + \frac{1}{2} \, ( 1-\delta_{ab} ) \left( \delta_{c 2} - \delta_{c 1} \right) \, \left( \lambda_3^{ab} + \lambda_4^{ab} \right) \, \frac{v^2}{m_{b}^{2} - m_{a}^{2}} \, ,
\end{equation}
which agrees with the expression given
in~\cite{Fuentes-Martin:2019dxt} if terms of order $s_{12}^2$ are
neglected.

\section{High-energy behavior}
\label{sec:running}

The conservation of the $\mathbb{Z}_2$ parity is crucial for the
Scotogenic setup to be consistent. In the absence of this symmetry,
neutrinos would acquire masses at tree-level and the DM candidate
would no longer be stable. This motivates the study of the
conservation of $\mathbb{Z}_2$ at high energies, a line of work
initiated in~\cite{Merle:2015gea}. As pointed out in this reference,
the RGE flow in the Scotogenic model might alter the shape of the
scalar potential at high energies and lead to the breaking of
$\mathbb{Z}_2$. This issue was fully explored in subsequent
works~\cite{Merle:2015ica,Lindner:2016kqk}, which show that the
breaking of the $\mathbb{Z}_2$ parity actually takes place in large
regions of the parameter space. A similar discussion for a variation
of the Scotogenic model including scalar and fermion triplets was
presented in~\cite{Merle:2016scw}.

Some general features of the high-energy behavior of the model, and in
particular of the possible breaking of the $\mathbb{Z}_2$ symmetry,
can be understood by inspecting the 1-loop $\beta$ function for the
$m_\eta^2$ parameter, shown in
Appendix~\ref{sec:app1}. Eq.~\eqref{eq:RGEmeta2} generalizes the
result previously derived in~\cite{Merle:2015gea} and gives the 1-loop
$\beta$ function for the $m_\eta^2$ matrix, valid for any values of
$(n_N,n_\eta)$. In order to study the possible breaking of
$\mathbb{Z}_2$, one must consider the sign (positive or negative) of
the individual contributions to the running of $m_\eta^2$. In this
regard, the negative contribution of the term proportional to $\Tr
\left[ y_a^\dagger M_N^\ast M_N y_b \right]$ turns out to be
crucial. In the following, we will refer to this term as \textit{the
  trace term}. As first pointed out in~\cite{Merle:2015gea} for the
standard Scotogenic model, in case of large Yukawa couplings
(equivalent to $\lambda_5 \ll 1$) and $M_N^2 \gtrsim m_\eta^2$, the
trace term dominates the $m_\eta^2$ running and drives it towards
negative values. Eventually, this leads to the breaking of the
$\mathbb{Z}_2$ symmetry at high energies, once $m_\eta^2 < 0$ induces
a minimum of the scalar potential with $\langle \eta \rangle \ne
0$. The same behavior is expected in the general Scotogenic
model. Other terms in Eq.~\eqref{eq:RGEmeta2} may counteract this
effect. In particular, the terms proportional to the quartic scalar
couplings may do so if their signs are properly chosen. The
contribution to the $m_\eta^2$ running will be positive for $\lambda_2
> 0$ and $\lambda_{3,4} < 0$ (since $m_H^2 < 0$), while their effect
will reinforce that of the trace term otherwise.

We will now explore the scalar potential of the model at high energies
by solving the full set of RGEs numerically. In order to do that we
will concentrate on two specific (but representative) versions of the
general Scotogenic model:
\begin{itemize}
\item The {\bf $\boldsymbol{(3,1)}$ model}, with three singlet
  fermions and one inert doublet. This is the original Scotogenic
  model~\cite{Ma:2006km}.
\item The {\bf $\boldsymbol{(1,3)}$ model}, with one singlet fermion
  and three inert doublets.
\end{itemize}
We set all model parameters at the electroweak scale, which we take to
be the $Z$-boson mass, $m_Z$. Therefore, in the following all values
for the input parameters must be understood to hold at $\mu = m_Z$. We
compute $m_H^2$ by solving the tadpole equations of the model and set
the $\lambda_1$ value to reproduce the measured Higgs boson mass. The
remaining scalar potential parameters are chosen freely, but always to
values that guarantee that the potential is bounded from below (BFB)
at the electroweak scale. This is a non-trivial requirement due to the
complexity of the scalar potential of the general Scotogenic model. We
refer to Appendix~\ref{sec:app2} for a detailed discussion on how we
check boundedness from below. Finally, we must accommodate the
neutrino squared mass differences and the leptonic mixing angles
measured in neutrino oscillation experiments by properly fixing the
Yukawa couplings of the model. In the two variants of the general
Scotogenic model considered the Yukawa couplings become $3 \times 3$
matrices, and then they can be obtained by means of a Casas-Ibarra
parametrization~\cite{Casas:2001sr}, adapted to the Scotogenic model
as explained
in~\cite{Toma:2013zsa,Vicente:2014wga,Cordero-Carrion:2018xre,Cordero-Carrion:2019qtu}. This
allows us to write the Yukawa matrices in full generality as
\begin{equation} \label{eq:CI}
y = i \, V^\dagger \, \Sigma^{-1/2} \, R \, D_{\sqrt{m}} \, U^\dagger \, .
\end{equation}
Here $U$ is a $3 \times 3$ unitary matrix, defined by the Takagi
decomposition of the neutrino mass matrix
\begin{equation}
U^T \, m_\nu \, U = \text{diag} \left( m_1, m_2, m_3 \right) \, ,
\end{equation}
with $m_i$ the three physical neutrino masses. $R$ is a general $3
\times 3$ orthogonal matrix and we have defined $D_{\sqrt{m}} =
\text{diag} \left( \sqrt{m_1}, \sqrt{m_2}, \sqrt{m_3}
\right)$. Finally, $\Sigma$ and $V$ are determined by the matrix $M$,
defined implicitly by the general expression $m_\nu = y^T \, M \, y$.
$\Sigma=\text{diag}\left(\sigma_1,\sigma_2,\sigma_3\right)$ is a
diagonal matrix containing the eigenvalues of $M$, while $V$ is a $3
\times 3$ unitary matrix such that $M = V^T \, \Sigma \, V$. Indeed,
as shown in Sec.~\ref{sec:numass}, the analytical expression for the
neutrino mass matrix in Eq.~\eqref{eq:mnu3} can be particularized to
the $(3,1)$ and $(1,3)$ models and in both cases one can write $m_\nu$
as the matrix product $y^T \, M \, y$, with different forms for the
matrix $M$. With these definitions, Eq.~\eqref{eq:CI} ensures
compatibility with neutrino oscillation data. We consider neutrino
normal mass ordering and the $1 \, \sigma$ ranges for the oscillation
parameters obtained in the global fit~\cite{deSalas:2017kay},
including the CP-violating phase $\delta$, hence allowing for complex
Yukawa couplings. For simplicity, we take $m_1 = 0$ and $R =
\mathbb{I}$, with $\mathbb{I}$ the $3 \times 3$ identity
matrix.~\footnote{For a general discussion on the parametrization of
  Yukawa couplings in Majorana neutrino mass models we refer
  to~\cite{Cordero-Carrion:2018xre,Cordero-Carrion:2019qtu}. Even
  though we have focused on the $(3,1)$ and $(1,3)$ Scotogenic models,
  in which the Yukawa couplings are matrices, we note that the
  \textit{master parametrization} introduced in these references can
  be used in variants of the general Scotogenic model with both $n_N,
  n_\eta > 1$, which can be regarded as \textit{hybrid scenarios}, see
  Appendix F of~\cite{Cordero-Carrion:2019qtu}.}

Some comments are in order before presenting our numerical results. In
what follows, several regions of the parameter spaces of the $(3,1)$
and $(1,3)$ Scotogenic models will be explored. Our focus is the study
of the behavior of these models at high energies. While several
phenomenological directions of interest can be pursued, these are
beyond the scope of our work. In particular, we are interested in
effects associated to the trace term, what motivates the consideration
of small $\lambda_5$ values ($\lambda_5^{aa} \leq 10^{-8}$). Larger
$\lambda_5$ entries would require smaller $y$ Yukawa couplings in
order to accommodate the mass scales measured in neutrino oscillations
experiments, see Eqs.~\eqref{eq:mnu31}, \eqref{eq:mnu12I} and
\eqref{eq:mnu12II}, hence making the trace term numerically less
relevant. For this reason, all scenarios considered below have $y \sim
\mathcal{O}(1)$. While this may lead to conflict with the current
bounds from the non-observation of charged lepton flavor violating
processes, we note the existence of many free parameters in the $y$
Yukawa matrices. This freedom can be used to cancel the most
constraining observables, for instance by choosing specific $R$
matrices, without any impact on our discussion. Similarly, the
scenarios considered below, and in particular the values chosen for
the masses of the $\mathbb{Z}_2$-odd states, may not be compatible
with the measured dark matter relic density.

\begin{figure}[tb!]
\centering
\includegraphics[width=0.49\linewidth,keepaspectratio]{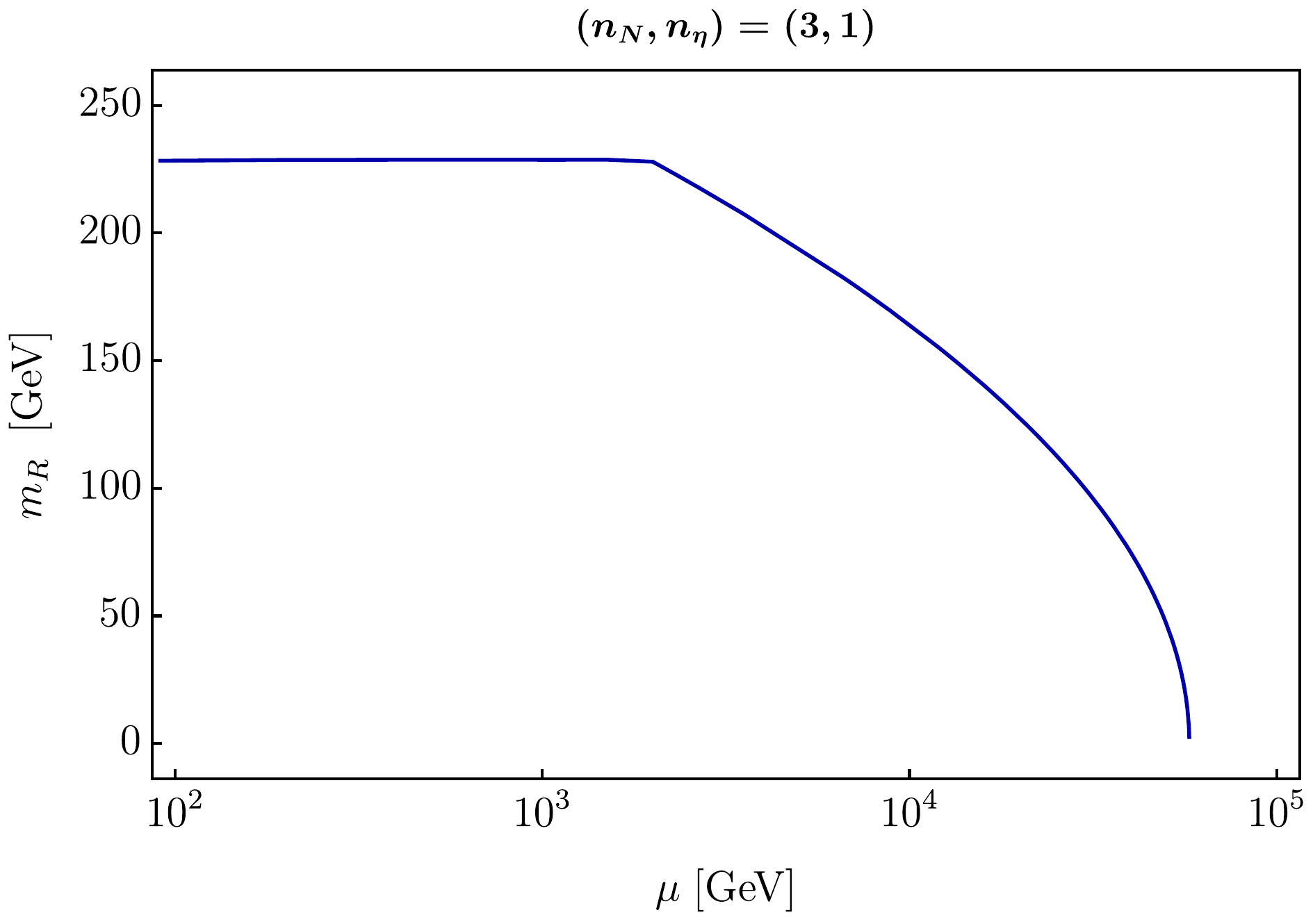}
\includegraphics[width=0.49\linewidth,keepaspectratio]{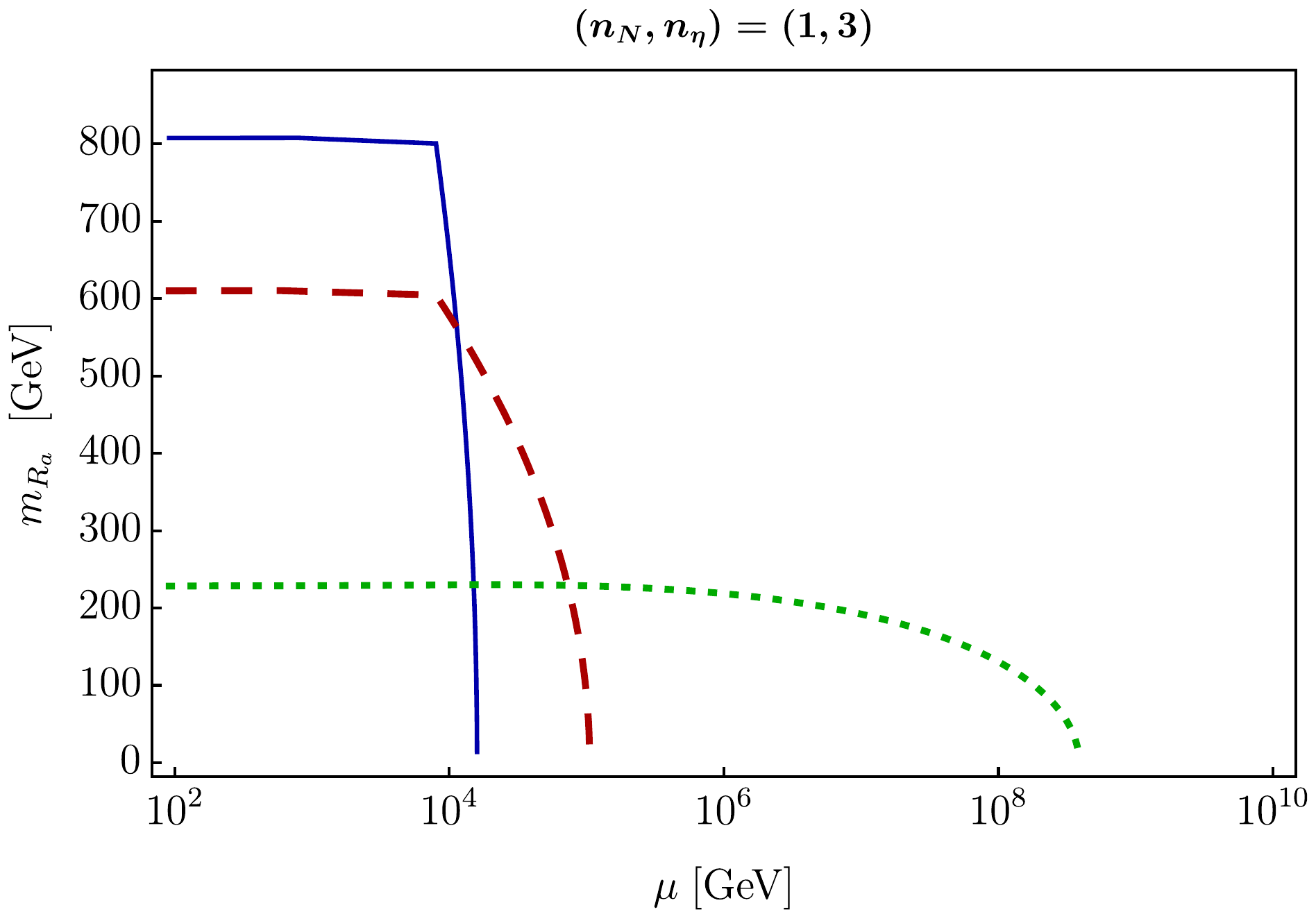}
\caption{Evolution of the CP-even scalar masses as a function of the
  energy scale $\mu$ in the $(3,1)$ and $(1,3)$ Scotogenic models. To
  the left, the CP-even scalar mass $m_R$ in the standard $(3,1)$
  model with $M_N = (1, 1.5, 2)$ TeV, $\lambda_2 = \lambda_3 =
  \lambda_4 = 0.1$, $\lambda_5 = 10^{-9}$ and $m_\eta^2 = (200$
  GeV$)^2$. To the right, the three CP-even scalar masses $m_{R_a}$ in
  the $(1,3)$ model with $M_N = 8$ TeV, $\lambda_2^{aaaa} =
  \lambda_3^{aa} = \lambda_4^{aa} = 0.1$, $\lambda_5^{aa} = 10^{-9}$
  and $m_\eta^2 = (200^2, 600^2, 800^2)$ GeV$^2$, with the remaining
  scalar parameters set to zero.
  \label{fig:Z2}}
\end{figure}

First, we have \textit{rediscovered} the parity problem in the
standard $(3,1)$ Scotogenic model. This is shown on the left-hand side
of Fig.~\ref{fig:Z2}, which displays the RGE evolution of the CP-even
scalar mass $m_R$ with the energy scale $\mu$. This is the most
convenient parameter to study the breaking of the $\mathbb{Z}_2$
symmetry. When $m_R^2$ becomes negative, the lightest CP-even scalar
becomes tachyonic, a clear sign that $\langle \eta \rangle = 0$ is not
the minimum of the potential. We have checked that the scalar
potential is BFB at all energy scales in this figure. We note that due
to our parameter choices the lightest singlet fermion, $N_1$, has
vanishing Yukawa couplings. For the same reason, $y_{2 \alpha} \ll
y_{3 \alpha}$ and the effect is driven predominantly by $N_3$. This
explains the drastic change in the evolution of $m_R$ at $\mu = 2$
TeV, when $N_3$ becomes active. Below this scale, $N_3$ effectively
decouples and does not contribute to the RGE running. We point out
that a much less pronounced change takes also place at $\mu =
1.5$~TeV, when $N_2$ becomes active, but this is not visible on the
figure. The $\mathbb{Z}_2$ parity gets broken at $\mu \simeq 60$~TeV,
after which the $\eta_R$ state becomes tachyonic. These results agree
well with those found in~\cite{Merle:2015gea} and confirm the possible
breaking of $\mathbb{Z}_2$ in the original Scotogenic model. A very
similar behavior is found for the $(1,3)$ model, which only has one
singlet fermion, as shown on the right-hand side of
Fig.~\ref{fig:Z2}. In this case, the three CP-even scalar masses
$m_{R_a}$ are displayed. Again, we have checked that the scalar
potential is BFB at all energy scales in this figure. As in the case
of the standard Scotogenic model, when one of the CP-even scalar
masses reaches zero the $\mathbb{Z}_2$ symmetry gets broken. We see in
this figure that this happens at $\mu \simeq 15$ TeV, where one of the
scalar masses (the one receiving the largest contribution from the
trace term) goes very sharply towards zero due to the effect of the
large $M_N = 8$ TeV value. This is clearly the same behavior observed
in the standard $(3,1)$ Scotogenic model.

\begin{figure}[tb!]
\centering
\includegraphics[width=0.65\linewidth,keepaspectratio]{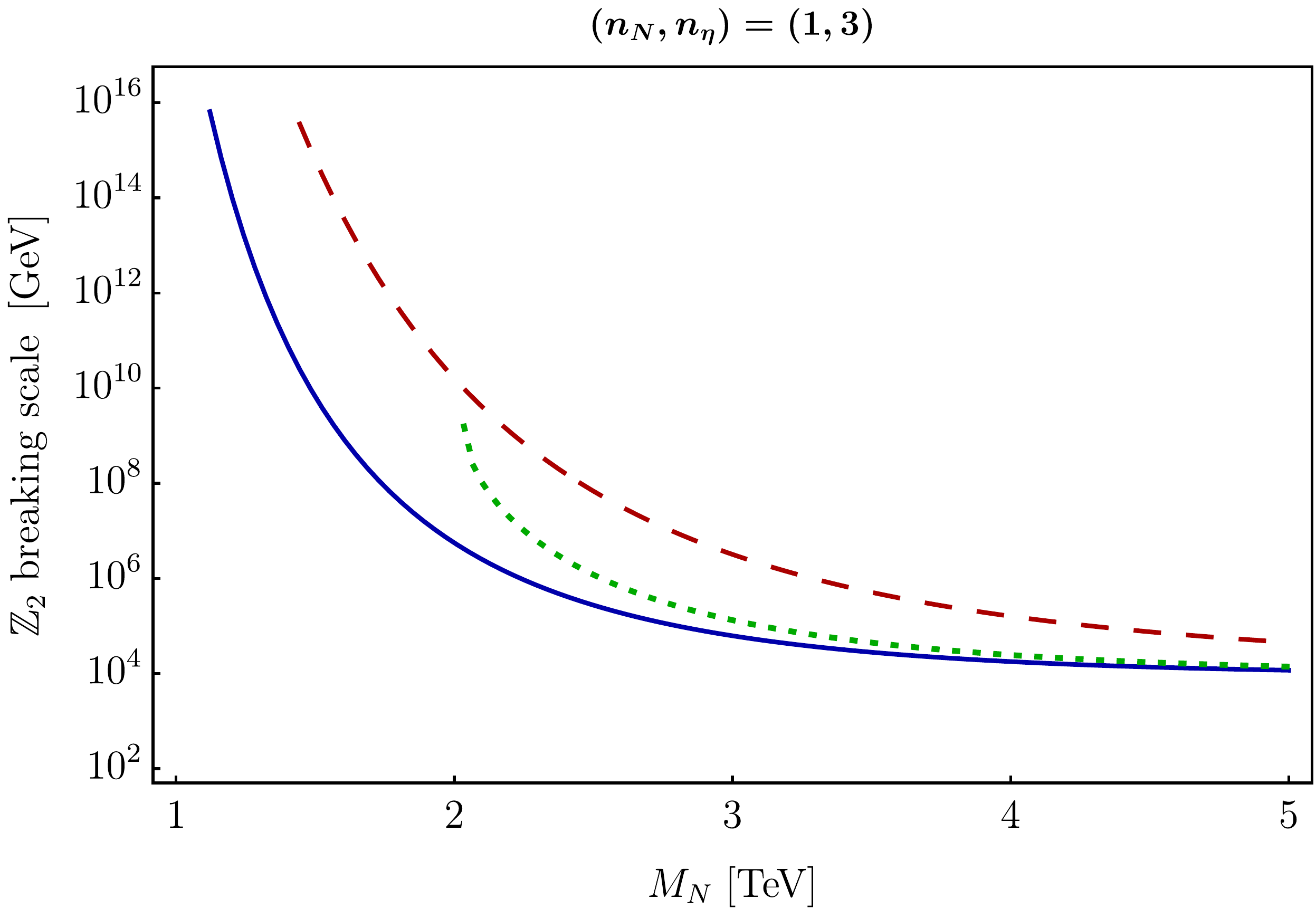}
\caption{$\mathbb{Z}_2$ breaking scale as a function of the singlet
  fermion mass $M_N$ in the $(1,3)$ Scotogenic model for three
  different scenarios: $\lambda_2^{aaaa} = \lambda_3^{aa} =
  \lambda_4^{aa} = 0.1$ and $m_\eta^2 = \left( 200^2, 300^2, 400^2
  \right)$~GeV$^2$ (blue), $\lambda_2^{aaaa} = \lambda_3^{aa} =
  \lambda_4^{aa} = 0.1$ and $m_\eta^2 = \left( 200^2, 600^2, 800^2
  \right)$~GeV$^2$ (red, dashed), and $\lambda_2^{aaaa} =
  \lambda_2^{aabb} = \lambda_3^{aa} = \lambda_4^{aa} = 0.3$ and
  $m_\eta^2 = \left( 200^2, 300^2, 400^2 \right)$~GeV$^2$ (green,
  dotted). In the three cases $\lambda_5^{aa} = 10^{-9}$ and the
  remaining quartic parameters are set to zero.
  \label{fig:Z2MN}}
\end{figure}

In the following we concentrate on the $(1,3)$ model. As already
discussed, the singlet fermion mass $M_N$ drives the scalar masses
towards negative values via the trace term, hence breaking the
$\mathbb{Z}_2$ parity at high energies. Fig.~\ref{fig:Z2MN} shows the
$\mathbb{Z}_2$ breaking scale as a function of $M_N$ for several
scalar parameter sets. The blue and red lines correspond to moderate
values for the quartic couplings, $\lambda_2^{aaaa} = \lambda_3^{aa} =
\lambda_4^{aa} = 0.1$, while the green line has increased (and
additional) quartics, $\lambda_2^{aaaa} = \lambda_2^{aabb} =
\lambda_3^{aa} = \lambda_4^{aa} = 0.3$. The $\lambda_5$ matrix is
taken to be diagonal, with $\lambda_5^{aa} = 10^{-9}$. We have
explicitly checked that the scalar potential is BFB at the electroweak
scale in all scenarios.~\footnote{We have allowed for (possible)
  non-BFB potentials at high energies, where some of the quartic
  couplings become negative due to running effects. We note that our
  algorithm gives us only sufficient (and not necessary) boundedness
  from below conditions, and in principle some of the \text{possibly}
  non-BFB potentials might actually be BFB. Morevoer, even non-BFB
  potentials may be realistic if the electroweak vacuum is metastable
  and has a large enough lifetime. This issue is already present in
  the SM and is clearly beyond the scope of our analysis, which
  focuses on the possible breaking of the $\mathbb{Z}_2$ symmetry.} As
expected, the $\mathbb{Z}_2$ breaking scale decreases for larger $M_N$
since the effect of the trace term becomes stronger. While different
scalar potential couplings may alter the outcome, this generic
behavior is found in large portions of the parameter space. One should
notice, however, that the green curve begins at $M_N \simeq 2$
TeV. For this specific scenario, lower values of $M_N$ do not break
the $\mathbb{Z}_2$ symmetry, as we now proceed to discuss.

\begin{figure}[tb!]
\centering
\includegraphics[width=0.65\linewidth,keepaspectratio]{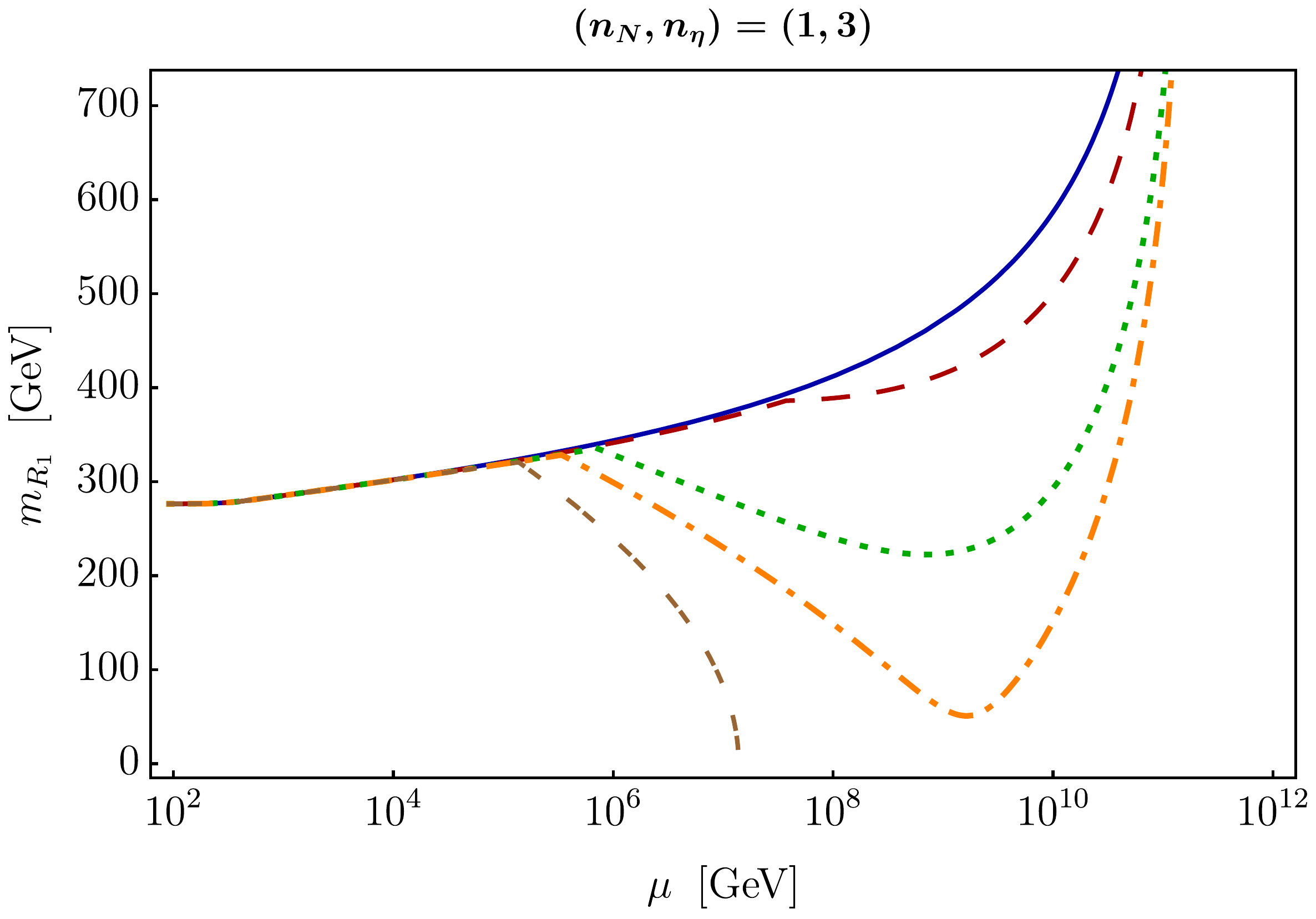}
\caption{Evolution of the lightest scalar mass $m_{R_1}$ as a function
  of the energy scale $\mu$ in the $(1,3)$ Scotogenic model. The
  scalar parameters are set to $\lambda_2^{aaaa} = \lambda_2^{aabb} =
  \lambda_3^{aa} = \lambda_4^{aa} = 0.3$, $\lambda_5^{aa} = 10^{-9}$
  and $m_\eta^2 = \left( 200^2, 300^2, 400^2 \right)$~GeV$^2$, while
  $M_N$ takes the values $1$~TeV (blue), $1.5$~TeV (red, dashed),
  $1.9$~TeV (green, dotted), $2.025$~TeV (orange, dash-dotted) and
  $2.2$~TeV (brown, double dashed).
  \label{fig:landau}}
\end{figure}

Fig.~\ref{fig:landau} shows the evolution of the lightest scalar mass
$m_{R_1}$ as a function of the energy for the parameter values
corresponding to the green curve in Fig.~\ref{fig:Z2MN}. The results
have been obtained for several values of $M_N$. It is important to
note that this figure shows the mass of the lightest scalar at each
energy scale, and not the mass of a single mass eigenstate at all
energies. For $M_N = 2.2$ TeV one observes that $m_{R_1}$ reaches zero
and the $\mathbb{Z}_2$ symmetry gets broken at $\mu \simeq 10^7$ GeV,
in accordance with Fig.~\ref{fig:Z2MN}. For lower $M_N$ values,
however, $m_{R_1}$ never reaches zero. Although $m_{R_1}$ gets
initially decreased due to the effect of the trace term, it eventually
increases at higher energies. The reason is the appearance of a Landau
pole in the $\lambda_2$ quartic couplings. In this figure
$\lambda_2^{aaaa} = \lambda_2^{aabb} = 0.3$ at the electroweak scale,
and this value grows with the energy until it completely dominates the
$m_\eta^2$ $\beta$ function with a positive contribution, see
Eq.~\eqref{eq:RGEmeta2}. The high multiplicity of $\lambda_2$
couplings reinforces the effect. Actually, we note that this Landau
pole is present at very high energies, well above the $\mathbb{Z}_2$
breaking scale, for many choices of the parameters at the electroweak
scale.

\begin{figure}[tb!]
\centering
\includegraphics[width=0.65\linewidth,keepaspectratio]{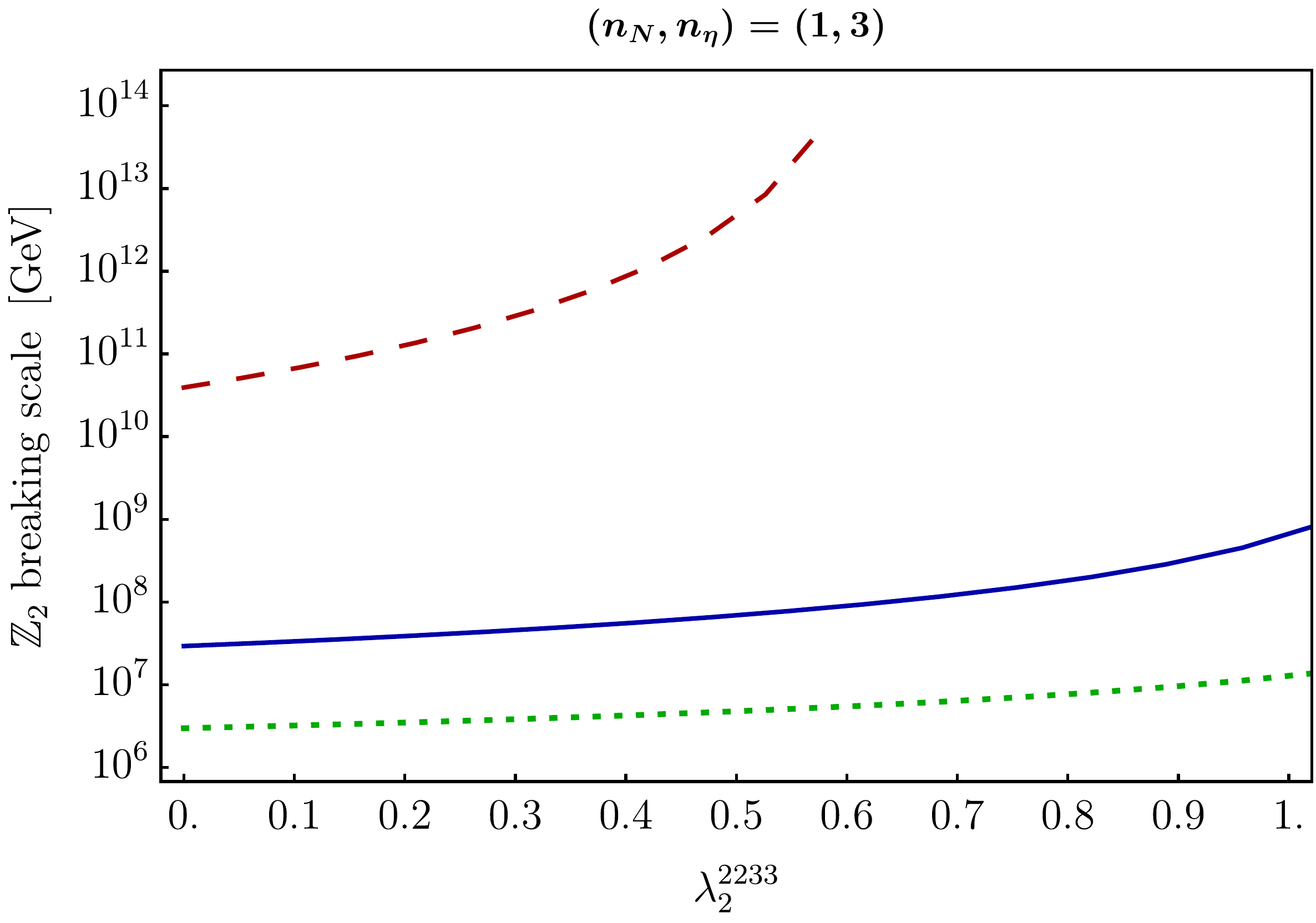}
\caption{$\mathbb{Z}_2$ breaking scale as a function of the
  $\lambda_2^{2233}$ parameter in the $(1,3)$ Scotogenic model for
  three different scenarios: $\lambda_2^{aaaa} = \lambda_3^{aa} =
  \lambda_4^{aa} = 0.1$, $\lambda_5^{aa} = 10^{-8}$, $m_\eta^2 =
  \left( 200^2, 300^2, 400^2 \right)$~GeV$^2$ and $M_N = 5$ TeV
  (blue), $\lambda_2^{aaaa} = \lambda_3^{aa} = \lambda_4^{aa} = 0.1$,
  $\lambda_5^{aa} = 10^{-9}$, $m_\eta^2 = \left( 200^2, 250^2, 300^2
  \right)$~GeV$^2$ and $M_N = 1.25$ TeV (red, dashed), and
  $\lambda_2^{aaaa} = \lambda_2^{aabb} = \lambda_3^{aa} =
  \lambda_4^{aa} = 0.3$, $\lambda_5^{aa} = 10^{-8}$, $m_\eta^2 =
  \left( 200^2, 600^2, 800^2 \right)$~GeV$^2$ and $M_N = 9$ TeV
  (green, dotted). In the three cases the remaining quartic parameters
  are set to zero.
  \label{fig:Z2lambda}}
\end{figure}

We conclude our exploration of the high-energy behavior of the $(1,3)$
model with Fig.~\ref{fig:Z2lambda}. In this case we plot the
$\mathbb{Z}_2$ breaking scale as a function of one of the $\lambda_2$
parameters, namely $\lambda_2^{2233}$. This is done for three
scenarios: the blue curve corresponds to $\lambda_2^{aaaa} =
\lambda_3^{aa} = \lambda_4^{aa} = 0.1$, $\lambda_5^{aa} = 10^{-8}$,
$m_\eta^2 = \left( 200^2, 300^2, 400^2 \right)$~GeV$^2$ and $M_N = 5$
TeV, in red we show the results for $\lambda_2^{aaaa} = \lambda_3^{aa}
= \lambda_4^{aa} = 0.1$, $\lambda_5^{aa} = 10^{-9}$, $m_\eta^2 =
\left( 200^2, 250^2, 300^2 \right)$~GeV$^2$ and $M_N = 1.25$ TeV,
while the green line is for $\lambda_2^{aaaa} = \lambda_2^{aabb} =
\lambda_3^{aa} = \lambda_4^{aa} = 0.3$, $\lambda_5^{aa} = 10^{-8}$,
$m_\eta^2 = \left( 200^2, 600^2, 800^2 \right)$~GeV$^2$ and $M_N = 9$
TeV. In all cases we have checked that the scalar potential is BFB at
the electroweak scale. For the blue and green lines, the impact of
$\lambda_2^{2233}$ is relatively mild. This is because the high values
of $M_N$ ($5$ and $9$ TeV, respectively) make the trace term
completely dominant and break the $\mathbb{Z}_2$ symmetry at
relatively low energies. In contrast, the $\mathbb{Z}_2$ breaking
scale has a much stronger dependence on $\lambda_2^{2233}$ in the red
scenario, which has a lower $M_N = 1.25$ TeV. For $\lambda_2^{2233}
\gtrsim 0.6$, a Landau pole is found \textit{before} the $\mathbb{Z}_2$
symmetry gets broken.

\section{Thermal effects and the fate of the $\boldsymbol{\mathbb{Z}_2}$ symmetry}
\label{sec:thermal}

To determine the cosmological impact of $\mathbb{Z}_2$ breaking one
needs to take into account thermal corrections. This is because the
interaction with the hot, primordial plasma induces an effective
potential for the scalar fields. This effective potential, at 1-loop
order, is given by
\begin{equation}
V_{\rm 1-loop}(\eta,T)=V_{\rm CW}(\eta)+\frac{T^4}{2\pi}\left[n_BJ_B(m^2(\eta)/T^2)-n_fJ_F(m_f(\eta)^2/T^2)\right]\,.
\end{equation}
Here $V_{\rm CW}$ is the standard Coleman-Weinberg potential for
$\eta$ at zero temperature while $J_B(m_b(\eta)^2/T^2)$ and
$J_F(m_f(\eta)^2/T^2)$ are the bosonic and fermionic functions,
respectively. These functions admit a high-T expansion
(see~\cite{Quiros:1999jp} for a review) which allows to write the
scalar mass as
\begin{equation}\label{themal_mass}
m_\eta^2(T)\sim m_\eta^2+c \, T^2\,.
\end{equation}
The coefficient $c$ depends on the details of the theory, such as the
quartic, gauge and Yukawa couplings.~\footnote{The thermal effects and
  phase transition have been extensively studied for the inert doublet
  model, see~\cite{Gil:2012ya,Blinov:2015vma}.} At any given time in
the early Universe, as it can be seen in Eq.~\eqref{themal_mass}, the
effect of the temperature is usually to restore the symmetry with the
subsequent \textit{dilution} of the effects of the running that we
have discussed in the previous section. It is therefore mandatory to
study if temperature has any impact on the fate of the $\mathbb{Z}_2$
symmetry and, therefore, on the stability of DM.

During inflation, the $\eta$ field is expected to have large quantum
fluctuations, comparable to the Hubble parameter in this period,
$H_I$. These fluctuations can be much larger than the scalar mass at
zero temperature and, acting as a sort of random walk, might bring the
field to a vacuum where the $\mathbb{Z}_2$ is broken. Right after
reheating, when the temperature is potentially very large, the thermal
mass of the scalar field may be large enough to overcome all breaking
effects. The reason is that, assuming the decay of the inflaton is
fast enough (instantaneous reheating, $\Gamma_\Phi\sim H_I$), the
reheating temperature is roughly given by~\cite{Linde:2005ht}
\begin{equation}
T_{\rm RH} \sim 10^{-1} \, \sqrt{H_I \, M_P} \,,
\end{equation}
where $M_P$ is the Planck mass. Note that this temperature is
generically much larger than $H_I$.
If the number of e-folds is not exceedingly large, $T_{\rm RH}$ is
expected to be larger than any field excursion and we expect
$m_\eta(T_{\rm RH})^2>0$. In addition, this also implies that
$m_\eta(T_{\rm RH})^2\sim c \, T_{\rm RH}^2\gg H(T_{\rm RH})^2$,
meaning that the field will fastly roll down to the minimum, at zero
value $\vev{\eta}=0$.~\footnote{In the thermal phase the field will
  experience oscillations around $\eta=0$ with an amplitude that
  decreases fast due to Hubble expansion and interactions with the
  thermal plasma.}

As the temperature decreases, it may happen that RGE effects make the
$\mathbb{Z}_2$ breaking to occur at some high-energy scale. However,
the $\eta$ field will be already at $\vev{\eta}=0$, meaning that it
cannot experience such a breaking. As the temperature continues
decreasing, we reach the freeze-out temperature. From this point on,
any breaking of the dark parity would be a disaster for the DM
stability. Note however, that since the $\eta$ field is at its local
minimum, $\vev{\eta}=0$, it cannot notice this high-energy RGE induced
symmetry breaking as it will only \textit{feel} the local properties
of the vacuum around $\vev{\eta}=0$.

Of course, this does not mean that RGE effects are completely harmless
for the Scotogenic model. In fact, the RGE-induced breaking could
induce the appearance of deeper minima in the potential, implying that
the stability of DM is just a local property of our vacuum, which
could be a \textit{false} vacuum, and not a global feature of the
potential.


\section{Summary and discussion}
\label{sec:conclusions}

The Scotogenic model is a well-known radiative scenario for the
generation of neutrino masses. The introduction of three singlet
fermions and one inert scalar doublet, all charged under a new
$\mathbb{Z}_2$ parity, leads to 1-loop Majorana neutrino masses and,
as a bonus, provides a viable weakly-interacting dark matter
candidate. In this work we have considered a generalization of this
setup to any numbers of generations of singlet fermions and inert
doublets. After computing the 1-loop neutrino mass matrix in the
general version of the model, we have studied its high-energy
behavior, focusing on two specific variants: the original Scotogenic
model with $(n_N,n_\eta) = (3,1)$ and a new multi-scalar variant with
$(n_N,n_\eta) = (1,3)$. Our main conclusion is that all the features
of the original model are kept in the multi-scalar version, with some
particularities due to the presence of a more involved scalar sector.

Our generalization of the Scotogenic model offer several novel
possibilities.  For instance, flavor model building could benefit from
an interesting feature of multi-scalar versions of the model. In the
$(n_N,n_\eta) = (1,3)$ model, one obtains three massive neutrinos and
leptonic mixing can be fully explained even if the Yukawa matrices are
completely diagonal. In this case the leptonic mixing matrix would be
generated by mixing in the scalar sector. This could be relevant in
some flavor models. For example, it may be a crucial ingredient to
rescue models where lepton mixing is predicted to be similar to quark
mixing. Novel phenomenological signatures might exist as well. The
$\eta$ doublets can be produced at the Large Hadron Collider due to
their couplings to the SM gauge bosons. Exotic signatures might be
possible in models with many $\eta$ generations, such as the
$(n_N,n_\eta) = (1,3)$ model. Cascade decays initiated by the
production of the heaviest $\eta$ doublets would lead to striking
multilepton signatures, including missing energy due to the production
of the lightest $\mathbb{Z}_2$-odd state at the end of the decay
chain. Finally, the dark matter production rates in the early Universe
might be affected as well by the presence of additional scalar degrees
of freedom. These interesting questions certainly deserve further
study.

\section*{Acknowledgements}

The authors are grateful to Igor Ivanov and Davide Racco for fruitful
discussions on the issues of boundedness from below and thermal
effects, respectively. Work supported by the Spanish grants
FPA2017-85216-P (MINECO/AEI/FEDER, UE), SEJI/2018/033 (Generalitat
Valenciana) and FPA2017-90566-REDC (Red Consolider MultiDark).  The
work of PE is supported by the FPI grant PRE2018-084599. The work of
MR is supported by the FPU grant FPU16/01907. AV acknowledges
financial support from MINECO through the Ramón y Cajal contract
RYC2018-025795-I.

\appendix

\section{Renormalization Group Equations}
\label{sec:app1}

At the 1-loop order, the RGEs of a model can be written as
\begin{equation}
\frac{d x(t)}{dt} = \frac{1}{16 \pi^2} \, \beta_x \, ,
\end{equation}
where $t \equiv \log \mu$, $\mu$ is the renormalization scale and
$\beta_x$ is the 1-loop $\beta$ function for the parameter $x$. In our
analysis, the full 1-loop running in the Scotogenic model with
arbitrary numbers of $N$ and $\eta$ generations has been
considered. Analytical expressions for all the 1-loop $\beta$
functions have been derived with the help of {\tt
  SARAH}~\cite{Staub:2008uz,Staub:2009bi,Staub:2010jh,Staub:2012pb,Staub:2013tta}.~\footnote{See~\cite{Vicente:2015zba}
  for a pedagogical introduction to the use of {\tt SARAH} in the
  context of non-supersymmetric models.} These have been included in a
code that solves the complete set of RGEs numerically.

We are mainly interested in the possible breaking of the
$\mathbb{Z}_2$ parity at high energies, and this is associated to the
running of the $m^2_\eta$ matrix. The corresponding 1-loop $\beta$
functions are given by
\begin{equation}
  \begin{split}
    \left(\beta_{m^2_\eta}\right)_{ab} = & - \frac{9}{10} \, g_1^2 \, \left(m^2_\eta\right)_{a b} - \frac{9}{2} \, g_2^2 \, \left(m^2_\eta\right)_{a b} + \sum_{c,d=1}^{\neta} \left[ 4 \, \lambda_2^{abcd} \left(m^2_\eta\right)_{dc} + 2 \, \lambda_2^{acdb} \left(m^2_\eta\right)_{cd} \right] \\
    & + \left[  4 \, \lambda_3^{ab} + 2 \, \lambda_4^{ab}  \right] \, m^2_H + \left(m^2_\eta\right)_{ab} \sum_{n = 1}^{\nN} \sum_{\alpha = 1}^3 \left( \left| y _{n a \alpha} \right|^2 +  \left| y _{n b \alpha} \right|^2  \right) - 4 \, \Tr \left[ y_a^\dagger M_N^\ast M_N y_b \right] \, .
  \end{split}
  \label{eq:RGEmeta2}
\end{equation}
Here $y_a \equiv \left[ y_{n a \alpha} \right]$ is a $\nN \times 3$ matrix,
being the first index a singlet fermion family index and the third one
a charged lepton family index. We have explicitly checked that for
$\nN=3$ and $\neta=1$, Eq.~\eqref{eq:RGEmeta2} reduces to the
$m^2_\eta$ $\beta$ function in the standard Scotogenic
model~\cite{Merle:2015gea}.

\section{Boundedness from below}
\label{sec:app2}

In order to ensure the existence of a stable vacuum, the scalar
potential of the theory must be BFB. There exist several approaches to
analyze boundedness from below. Ideally, one would like to have a
\textit{BFB test} that provides necessary and sufficient
conditions. This way, one could not only guarantee that all potentials
that pass the test are BFB (sufficient condition), but also discard
potentials that fail it (necessary condition). In this regard, a major step forward was given in~\cite{Kannike:2016fmd} and more recently in~\cite{Ivanov:2018jmz}. The algorithm proposed in the second reference provides necessary and sufficient conditions for boundedness from below in a generic scalar potential using notions of spectral theory of tensors. However, applying this
algorithm beyond a few simple cases turns out to be impractical due to
the computational cost involved. For this reason, in phenomenological
analyses one usually resorts to less ambitious approaches which only
provide sufficient conditions, but not necessary. These methods are
overconstraining, since one must reject potentials not passing the
test, even though they might actually be BFB. Nevertheless, if the
potential passes the test, one can fully trust that boundedness from
below is guaranteed.

Here we will employ the \textit{copositivity} criterion, which
combined with a recently developed mathematical algorithm, never
applied to a high-energy physics scenario, will give us sufficient
(but not necessary) conditions.  To the best of our knowledge, the
first paper relating copositivity with boundedness from below
was~\cite{Kannike:2012pe}. One must first express the quartic part of
the scalar potential, $\mathcal{V}_4$, as a quadratic form of the $n$
real fields $\varphi_a$ ($a = 1, 2, \dots n$) in the theory,
\begin{equation}
  \mathcal{V}_4 = \Lambda_{ab} \, \varphi_a^2 \varphi_b^2 \, .
\end{equation}
The scalar potential is BFB if and only if the matrix of quartic
couplings $\Lambda_{ab}$ is copositive. A real matrix $A$ is said to
be copositive if $x^T A \, x \geqslant 0$ for every non-negative
vector $x \geqslant 0$, that is, $x^i \geqslant 0$. If the inequality
is strict, the matrix is \emph{strictly copositive}. Therefore,
checking for the copositivity of the matrix of quartic couplings would
in principle provide sufficient and necessary boundedness from below
conditions. However, in complicated models such as the general
Scotogenic model, one cannot write $\mathcal{V}_4$ as a quadratic form
without introducing mixed bilinears (scalar field combinations
involving two different fields). For this reason, this method only
leads to sufficient conditions, as we now explain.

In order to write the quartic part of the scalar potential as a
quadratic form we define
\begin{equation}
  \varphi_i^\dagger \varphi_i = h_i^2 \, , \qquad \varphi_i^\dagger \varphi_j = \left|h_i \right| \left| h_j \right| \rho_{ij} e^{i \phi_{ij}} = h_{ij}^2 \, \rho_{ij} e^{i \phi_{ij}} \, \, ,
  \label{eq:bilinears}
\end{equation}
with $\left| \rho_{ij} \right| \in [0,1]$ by virtue of the
Cauchy-Schwarz inequality. Thus, we can express the boundedness from
below condition as
\begin{equation}
  \mathcal{V}_{4} = x^T \, V_{4} \, x \geqslant 0 \, ,
\end{equation}
with $x = \left( h_1^2 \ \dots \ h_i^2 \ \dots \ h_{ij}^2 \ \dots
\right)$ and the matrix $V_4$ is given by a combination of the quartic
couplings, the $\lambda$'s, as well as the $\rho$'s and $\phi$
phases.~\footnote{In the model under consideration, this includes also
  the phases of the $\lambda_5$ couplings.} The reason why this method
provides only sufficient conditions is the presence of the mixed
bilinears. Notice that the \textit{direction} given by $h_{ij}^2$ is
not independent of $h_i^2$ and $h_j^2$. Therefore, imposing $x^T \,
V_{4} \, x \geqslant 0$ for every non-negative $x$ vector is
overconstraining, since unphysical directions would be included in the
test. Nonetheless, when the test is positive, the potential is BFB. In
summary, a scalar potential is BFB if the associated $V_4$ matrix is
copositive. However, when the matrix is not copositive nothing can be
said about the potential.

There is mathematical work showing that a symmetric matrix $A$ of
order $n$ is (strictly) copositive if and only if every principal
submatrix $B$ of $A$ has no eigenvector $w > 0$ with associated
eigenvalue $\kappa < 0 \ (\leqslant 0)$~\cite{KAPLAN2000203}. However,
these theorems are of little practical value when the matrix has a
large order, since there will be $2^n - 1$ principal
submatrices. Luckily, we can make use of~\cite{Yang-Li} instead. The
authors of this work proposed an algorithm that leads to necessary and
sufficient conditions for the copositivity of unit diagonal matrices
(matrices with all diagonal elements equal to $1$). Although the
algorithm in~\cite{Yang-Li} could only be applied for up to $7 \times
7$ matrices, incidentally the case in the $(1,3)$ Scotogenic model,
more recent work by the same authors contains indications to extend it
to higher orders~\cite{Yang2010}.

After all these considerations, our procedure to check for
copositivity is as follows:

\begin{enumerate}
\item We replace all the quartic couplings in $V_4$ by the numerical
  values in the scalar potential we want to test.
\item We transform each element of the matrix to the \textit{worst
  case scenario}. This is achieved by treating the remaining $\rho$
  and $\phi$ parameters as independent variables and setting them to
  the configuration for which the term is minimal.~\footnote{We
    emphasize that we do this for each element. This means that even
    if the same $\rho$ parameter appears in two elements, it is
    treated as if each appearance is independent. This way we make
    sure that all the negative directions in the scalar potential are
    considered. However, we are again taking an overconstraining (and
    then very conservative) approach.\label{foot}}
\item We check if the matrix has null entries in the diagonal. If it
  does, we remove the corresponding rows and columns. The original
  matrix will be copositive if the remaining one is and the removed
  elements are non-negative.
\item We need the matrix to have unit diagonal to be able to apply the
  algorithm in~\cite{Yang-Li}. Therefore, we divide all its entries by
  the smallest element in the diagonal and we replace all the values
  greater than $1$ by $1$. The original matrix will be copositive if
  the new one is.
  \item We finally check the copositivity of the resulting matrix with
    the algorithm in~\cite{Yang-Li}.
\end{enumerate}

A final remark about our method is in order. The stability in
charge-breaking directions is ignored in many analyses. However, since
we are being overly restrictive treating all the $\rho$ moduli and
$\phi$ phases as independent variables in the different entries of
$V_4$, charge-breaking directions are included as well in our BFB
test. In order to prove it, let us parametrize the scalar doublets of
the model under consideration as
\begin{equation}
	\phi_{i}=\sqrt{r_{i}} e^{i \gamma_{i}}\left(\begin{array}{c}
	\sin \left(\alpha_{i}\right) \\
	\cos \left(\alpha_{i}\right) e^{i \beta_{i}}
	\end{array}\right) \, .
\end{equation}
This parametrization and an example of how to use it to explore
boundedness from below is shown in~\cite{Faro:2019vcd}.  Let us
consider a contraction of scalar doublets
\begin{equation}
  \left( \phi_i^\dagger \phi_j \right) = \sqrt{ r_{i} r_{j} } \left[ \sin \alpha_{i} \sin \alpha_{j} + \cos \alpha_{i} \cos \alpha_{j} e^{- i \left( \beta_{i} - \beta_{j} \right)} \right] \, ,
\end{equation}
and take the modulus of the term in square brackets
\begin{equation}  
  \begin{array}{l}
    \left|\sin \alpha_{i} \sin \alpha_{j}+\cos \alpha_{i} \cos \alpha_{j} e^{i \beta}\right|^{2} \\
    \quad=\sin ^{2} \alpha_{i} \sin ^{2} \alpha_{j}+\cos ^{2} \alpha_{i} \cos ^{2} \alpha_{j}+\sin \alpha_{i} \sin \alpha_{j} \cos \alpha_{i} \cos \alpha_{j}\left(e^{i \beta}+e^{-i \beta}\right) \\
    \quad=\sin ^{2} \alpha_{i} \sin ^{2} \alpha_{j}+\cos ^{2} \alpha_{i} \cos ^{2} \alpha_{j}+2 \sin \alpha_{i} \sin \alpha_{j} \cos \alpha_{i} \cos \alpha_{j} \cos \beta \leqslant 1 .
  \end{array}
\end{equation}
As expected, the product is, at most, as large as the modulus of the
fields, $\sqrt{r_i}$. Therefore, if we treat the factors that multiply
$\sqrt{r_i r_j}$ as independent variables (that is, being overly
restrictive as explained in footnote \ref{foot}), $\rho_{ij} e^{i
  \phi_{ij}}$, and make all combinations minimal, our method will
cover boundedness from below in charge-breaking directions as well.

\bibliographystyle{utphys}
\bibliography{refs}

\providecommand{\href}[2]{#2}\begingroup\raggedright\begin{thebibliography}{100}

\bibitem{Zee:1980ai}
A.~Zee, ``{A Theory of Lepton Number Violation, Neutrino Majorana Mass, and
  Oscillation},'' \href{http://dx.doi.org/10.1016/0370-2693(80)90349-4,
  10.1016/0370-2693(80)90193-8}{{\em Phys. Lett.} {\bfseries 93B} (1980) 389}.
[Erratum: Phys. Lett.95B,461(1980)].

\bibitem{Cheng:1980qt}
T.~P. Cheng and L.-F. Li, ``{Neutrino Masses, Mixings and Oscillations in $\rm
  SU(2) \times U(1)$ Models of Electroweak Interactions},''
\href{http://dx.doi.org/10.1103/PhysRevD.22.2860}{{\em Phys. Rev.} {\bfseries
  D22} (1980) 2860}.

\bibitem{Zee:1985id}
A.~Zee, ``{Quantum Numbers of Majorana Neutrino Masses},''
\href{http://dx.doi.org/10.1016/0550-3213(86)90475-X}{{\em Nucl. Phys.}
  {\bfseries B264} (1986) 99--110}.

\bibitem{Babu:1988ki}
K.~S. Babu, ``{Model of 'Calculable' Majorana Neutrino Masses},''
\href{http://dx.doi.org/10.1016/0370-2693(88)91584-5}{{\em Phys. Lett.}
  {\bfseries B203} (1988) 132--136}.

\bibitem{Cai:2017jrq}
Y.~Cai, J.~Herrero-Garc\'ia, M.~A. Schmidt, A.~Vicente, and R.~R. Volkas,
  ``{From the trees to the forest: a review of radiative neutrino mass
  models},'' \href{http://dx.doi.org/10.3389/fphy.2017.00063}{{\em Front.in
  Phys.} {\bfseries 5} (2017) 63},
\href{http://arxiv.org/abs/1706.08524}{{\ttfamily arXiv:1706.08524 [hep-ph]}}.

\bibitem{Restrepo:2013aga}
D.~Restrepo, O.~Zapata, and C.~E. Yaguna, ``{Models with radiative neutrino
  masses and viable dark matter candidates},''
  \href{http://dx.doi.org/10.1007/JHEP11(2013)011}{{\em JHEP} {\bfseries 11}
  (2013) 011},
\href{http://arxiv.org/abs/1308.3655}{{\ttfamily arXiv:1308.3655 [hep-ph]}}.

\bibitem{Ma:2006km}
E.~Ma, ``{Verifiable radiative seesaw mechanism of neutrino mass and dark
  matter},'' \href{http://dx.doi.org/10.1103/PhysRevD.73.077301}{{\em Phys.
  Rev.} {\bfseries D73} (2006) 077301},
\href{http://arxiv.org/abs/hep-ph/0601225}{{\ttfamily arXiv:hep-ph/0601225
  [hep-ph]}}.

\bibitem{FileviezPerez:2009ud}
P.~Fileviez~Perez and M.~B. Wise, ``{On the Origin of Neutrino Masses},''
  \href{http://dx.doi.org/10.1103/PhysRevD.80.053006}{{\em Phys. Rev.}
  {\bfseries D80} (2009) 053006},
\href{http://arxiv.org/abs/0906.2950}{{\ttfamily arXiv:0906.2950 [hep-ph]}}.

\bibitem{Liao:2009fm}
Y.~Liao and J.-Y. Liu, ``{Radiative and flavor-violating transitions of leptons
  from interactions with color-octet particles},''
  \href{http://dx.doi.org/10.1103/PhysRevD.81.013004}{{\em Phys. Rev.}
  {\bfseries D81} (2010) 013004},
\href{http://arxiv.org/abs/0911.3711}{{\ttfamily arXiv:0911.3711 [hep-ph]}}.

\bibitem{Reig:2018mdk}
M.~Reig, D.~Restrepo, J.~W.~F. Valle, and O.~Zapata, ``{Bound-state dark matter
  and Dirac neutrino masses},''
  \href{http://dx.doi.org/10.1103/PhysRevD.97.115032}{{\em Phys. Rev.}
  {\bfseries D97} no.~11, (2018) 115032},
\href{http://arxiv.org/abs/1803.08528}{{\ttfamily arXiv:1803.08528 [hep-ph]}}.

\bibitem{Reig:2018ztc}
M.~Reig, D.~Restrepo, J.~W.~F. Valle, and O.~Zapata, ``{Bound-state dark matter
  with Majorana neutrinos},''
  \href{http://dx.doi.org/10.1016/j.physletb.2019.01.023}{{\em Phys. Lett.}
  {\bfseries B790} (2019) 303--307},
\href{http://arxiv.org/abs/1806.09977}{{\ttfamily arXiv:1806.09977 [hep-ph]}}.

\bibitem{Farzan:2012sa}
Y.~Farzan and E.~Ma, ``{Dirac neutrino mass generation from dark matter},''
  \href{http://dx.doi.org/10.1103/PhysRevD.86.033007}{{\em Phys. Rev.}
  {\bfseries D86} (2012) 033007},
\href{http://arxiv.org/abs/1204.4890}{{\ttfamily arXiv:1204.4890 [hep-ph]}}.

\bibitem{Wang:2017mcy}
W.~Wang, R.~Wang, Z.-L. Han, and J.-Z. Han, ``{The $B-L$ Scotogenic Models for
  Dirac Neutrino Masses},''
  \href{http://dx.doi.org/10.1140/epjc/s10052-017-5446-9}{{\em Eur. Phys. J.}
  {\bfseries C77} no.~12, (2017) 889},
\href{http://arxiv.org/abs/1705.00414}{{\ttfamily arXiv:1705.00414 [hep-ph]}}.

\bibitem{Han:2018zcn}
Z.-L. Han and W.~Wang, ``{$Z'$ Portal Dark Matter in $B-L$ Scotogenic Dirac
  Model},'' \href{http://dx.doi.org/10.1140/epjc/s10052-018-6308-9}{{\em Eur.
  Phys. J.} {\bfseries C78} no.~10, (2018) 839},
\href{http://arxiv.org/abs/1805.02025}{{\ttfamily arXiv:1805.02025 [hep-ph]}}.

\bibitem{Calle:2018ovc}
J.~Calle, D.~Restrepo, C.~E. Yaguna, and O.~Zapata, ``{Minimal radiative Dirac
  neutrino mass models},''
  \href{http://dx.doi.org/10.1103/PhysRevD.99.075008}{{\em Phys. Rev.}
  {\bfseries D99} no.~7, (2019) 075008},
\href{http://arxiv.org/abs/1812.05523}{{\ttfamily arXiv:1812.05523 [hep-ph]}}.

\bibitem{Ma:2019yfo}
E.~Ma, ``{Scotogenic $U(1)_\chi$ Dirac neutrinos},''
  \href{http://dx.doi.org/10.1016/j.physletb.2019.05.006}{{\em Phys. Lett.}
  {\bfseries B793} (2019) 411--414},
\href{http://arxiv.org/abs/1901.09091}{{\ttfamily arXiv:1901.09091 [hep-ph]}}.

\bibitem{Ma:2019iwj}
E.~Ma, ``{Scotogenic cobimaximal Dirac neutrino mixing from $\Delta (27)$ and
  $U(1)_\chi $},'' \href{http://dx.doi.org/10.1140/epjc/s10052-019-7440-x}{{\em
  Eur. Phys. J.} {\bfseries C79} no.~11, (2019) 903},
\href{http://arxiv.org/abs/1905.01535}{{\ttfamily arXiv:1905.01535 [hep-ph]}}.

\bibitem{CentellesChulia:2019gic}
S.~Centelles~Chuli\'a, R.~Cepedello, E.~Peinado, and R.~Srivastava,
  ``{Scotogenic Dark Symmetry as a residual subgroup of Standard Model
  Symmetries},''
\href{http://arxiv.org/abs/1901.06402}{{\ttfamily arXiv:1901.06402 [hep-ph]}}.

\bibitem{Jana:2019mez}
S.~Jana, P.~Vishnu, and S.~Saad, ``{Minimal dirac neutrino mass models from
  $\hbox {U}(1)_{\mathrm{R}}$ gauge symmetry and left--right asymmetry at
  colliders},'' \href{http://dx.doi.org/10.1140/epjc/s10052-019-7441-9}{{\em
  Eur. Phys. J. C} {\bfseries 79} no.~11, (2019) 916},
  \href{http://arxiv.org/abs/1904.07407}{{\ttfamily arXiv:1904.07407
  [hep-ph]}}.

\bibitem{Jana:2019mgj}
S.~Jana, P.~Vishnu, and S.~Saad, ``{Minimal Realizations of Dirac Neutrino Mass
  from Generic One-loop and Two-loop Topologies at $d=5$},''
  \href{http://dx.doi.org/10.1088/1475-7516/2020/04/018}{{\em JCAP} {\bfseries
  04} (2020) 018}, \href{http://arxiv.org/abs/1910.09537}{{\ttfamily
  arXiv:1910.09537 [hep-ph]}}.

\bibitem{Ma:2008cu}
E.~Ma and D.~Suematsu, ``{Fermion Triplet Dark Matter and Radiative Neutrino
  Mass},'' \href{http://dx.doi.org/10.1142/S021773230903059X}{{\em Mod. Phys.
  Lett.} {\bfseries A24} (2009) 583--589},
\href{http://arxiv.org/abs/0809.0942}{{\ttfamily arXiv:0809.0942 [hep-ph]}}.

\bibitem{Ma:2008ym}
E.~Ma, ``{Dark Scalar Doublets and Neutrino Tribimaximal Mixing from A(4)
  Symmetry},'' \href{http://dx.doi.org/10.1016/j.physletb.2008.12.038}{{\em
  Phys. Lett.} {\bfseries B671} (2009) 366--368},
\href{http://arxiv.org/abs/0808.1729}{{\ttfamily arXiv:0808.1729 [hep-ph]}}.

\bibitem{Farzan:2009ji}
Y.~Farzan, ``{A Minimal model linking two great mysteries: neutrino mass and
  dark matter},'' \href{http://dx.doi.org/10.1103/PhysRevD.80.073009}{{\em
  Phys. Rev.} {\bfseries D80} (2009) 073009},
\href{http://arxiv.org/abs/0908.3729}{{\ttfamily arXiv:0908.3729 [hep-ph]}}.

\bibitem{Chen:2009gd}
C.-H. Chen, C.-Q. Geng, and D.~V. Zhuridov, ``{Neutrino Masses, Leptogenesis
  and Decaying Dark Matter},''
  \href{http://dx.doi.org/10.1088/1475-7516/2009/10/001}{{\em JCAP} {\bfseries
  0910} (2009) 001},
\href{http://arxiv.org/abs/0906.1646}{{\ttfamily arXiv:0906.1646 [hep-ph]}}.

\bibitem{Adulpravitchai:2009re}
A.~Adulpravitchai, M.~Lindner, A.~Merle, and R.~N. Mohapatra, ``{Radiative
  Transmission of Lepton Flavor Hierarchies},''
  \href{http://dx.doi.org/10.1016/j.physletb.2009.09.042}{{\em Phys. Lett.}
  {\bfseries B680} (2009) 476--479},
\href{http://arxiv.org/abs/0908.0470}{{\ttfamily arXiv:0908.0470 [hep-ph]}}.

\bibitem{Farzan:2010mr}
Y.~Farzan, S.~Pascoli, and M.~A. Schmidt, ``{AMEND: A model explaining neutrino
  masses and dark matter testable at the LHC and MEG},''
  \href{http://dx.doi.org/10.1007/JHEP10(2010)111}{{\em JHEP} {\bfseries 10}
  (2010) 111},
\href{http://arxiv.org/abs/1005.5323}{{\ttfamily arXiv:1005.5323 [hep-ph]}}.

\bibitem{Aoki:2011yk}
M.~Aoki, S.~Kanemura, and K.~Yagyu, ``{Doubly-charged scalar bosons from the
  doublet},'' \href{http://dx.doi.org/10.1016/j.physletb.2011.11.043,
  10.1016/j.physletb.2011.07.017}{{\em Phys. Lett.} {\bfseries B702} (2011)
  355--358}, \href{http://arxiv.org/abs/1105.2075}{{\ttfamily arXiv:1105.2075
  [hep-ph]}}.
[Erratum: Phys. Lett.B706,495(2012)].

\bibitem{Cai:2011qr}
Y.~Cai, X.-G. He, M.~Ramsey-Musolf, and L.-H. Tsai, ``{R$\nu$MDM and Lepton
  Flavor Violation},'' \href{http://dx.doi.org/10.1007/JHEP12(2011)054}{{\em
  JHEP} {\bfseries 12} (2011) 054},
\href{http://arxiv.org/abs/1108.0969}{{\ttfamily arXiv:1108.0969 [hep-ph]}}.

\bibitem{Chen:2011bc}
C.-H. Chen and S.~S.~C. Law, ``{Exotic fermion multiplets as a solution to
  baryon asymmetry, dark matter and neutrino masses},''
  \href{http://dx.doi.org/10.1103/PhysRevD.85.055012}{{\em Phys. Rev.}
  {\bfseries D85} (2012) 055012},
\href{http://arxiv.org/abs/1111.5462}{{\ttfamily arXiv:1111.5462 [hep-ph]}}.

\bibitem{Chao:2012sz}
W.~Chao, ``{Dark matter, LFV and neutrino magnetic moment in the radiative
  seesaw model with fermion triplet},''
  \href{http://dx.doi.org/10.1142/S0217751X15500074}{{\em Int. J. Mod. Phys.}
  {\bfseries A30} no.~01, (2015) 1550007},
\href{http://arxiv.org/abs/1202.6394}{{\ttfamily arXiv:1202.6394 [hep-ph]}}.

\bibitem{Ma:2012ez}
E.~Ma, A.~Natale, and A.~Rashed, ``{Scotogenic $A_4$ Neutrino Model for Nonzero
  $\theta_{13}$ and Large $\delta_{CP}$},''
  \href{http://dx.doi.org/10.1142/S0217751X12501345}{{\em Int. J. Mod. Phys.}
  {\bfseries A27} (2012) 1250134},
\href{http://arxiv.org/abs/1206.1570}{{\ttfamily arXiv:1206.1570 [hep-ph]}}.

\bibitem{Hirsch:2013ola}
M.~Hirsch, R.~A. Lineros, S.~Morisi, J.~Palacio, N.~Rojas, and J.~W.~F. Valle,
  ``{WIMP dark matter as radiative neutrino mass messenger},''
  \href{http://dx.doi.org/10.1007/JHEP10(2013)149}{{\em JHEP} {\bfseries 10}
  (2013) 149},
\href{http://arxiv.org/abs/1307.8134}{{\ttfamily arXiv:1307.8134 [hep-ph]}}.

\bibitem{Bhattacharya:2013mpa}
S.~Bhattacharya, E.~Ma, A.~Natale, and A.~Rashed, ``{Radiative Scaling Neutrino
  Mass with $A_4$ Symmetry},''
  \href{http://dx.doi.org/10.1103/PhysRevD.87.097301}{{\em Phys. Rev.}
  {\bfseries D87} (2013) 097301},
\href{http://arxiv.org/abs/1302.6266}{{\ttfamily arXiv:1302.6266 [hep-ph]}}.

\bibitem{Ma:2013xqa}
E.~Ma, ``{Neutrino Mixing and Geometric CP Violation with Delta(27)
  Symmetry},'' \href{http://dx.doi.org/10.1016/j.physletb.2013.05.011}{{\em
  Phys. Lett.} {\bfseries B723} (2013) 161--163},
\href{http://arxiv.org/abs/1304.1603}{{\ttfamily arXiv:1304.1603 [hep-ph]}}.

\bibitem{Ma:2013nga}
E.~Ma, ``{Unified Framework for Matter, Dark Matter, and Radiative Neutrino
  Mass},'' \href{http://dx.doi.org/10.1103/PhysRevD.88.117702}{{\em Phys. Rev.}
  {\bfseries D88} no.~11, (2013) 117702},
\href{http://arxiv.org/abs/1307.7064}{{\ttfamily arXiv:1307.7064 [hep-ph]}}.

\bibitem{Brdar:2013iea}
V.~Brdar, I.~Picek, and B.~Radovcic, ``{Radiative Neutrino Mass with Scotogenic
  Scalar Triplet},''
  \href{http://dx.doi.org/10.1016/j.physletb.2013.11.045}{{\em Phys. Lett.}
  {\bfseries B728} (2014) 198--201},
\href{http://arxiv.org/abs/1310.3183}{{\ttfamily arXiv:1310.3183 [hep-ph]}}.

\bibitem{Law:2013saa}
S.~S.~C. Law and K.~L. McDonald, ``{A Class of Inert N-tuplet Models with
  Radiative Neutrino Mass and Dark Matter},''
  \href{http://dx.doi.org/10.1007/JHEP09(2013)092}{{\em JHEP} {\bfseries 09}
  (2013) 092},
\href{http://arxiv.org/abs/1305.6467}{{\ttfamily arXiv:1305.6467 [hep-ph]}}.

\bibitem{Patra:2014sua}
S.~Patra, N.~Sahoo, and N.~Sahu, ``{Dipolar dark matter in light of the 3.5 keV
  x-ray line, neutrino mass, and LUX data},''
  \href{http://dx.doi.org/10.1103/PhysRevD.91.115013}{{\em Phys. Rev.}
  {\bfseries D91} no.~11, (2015) 115013},
\href{http://arxiv.org/abs/1412.4253}{{\ttfamily arXiv:1412.4253 [hep-ph]}}.

\bibitem{Ma:2014eka}
E.~Ma and A.~Natale, ``{Scotogenic $Z_2$ or $U(1)_D$ Model of Neutrino Mass
  with $\Delta(27)$ Symmetry},''
  \href{http://dx.doi.org/10.1016/j.physletb.2014.05.070}{{\em Phys. Lett.}
  {\bfseries B734} (2014) 403--405},
\href{http://arxiv.org/abs/1403.6772}{{\ttfamily arXiv:1403.6772 [hep-ph]}}.

\bibitem{Fraser:2014yha}
S.~Fraser, E.~Ma, and O.~Popov, ``{Scotogenic Inverse Seesaw Model of Neutrino
  Mass},'' \href{http://dx.doi.org/10.1016/j.physletb.2014.08.069}{{\em Phys.
  Lett.} {\bfseries B737} (2014) 280--282},
\href{http://arxiv.org/abs/1408.4785}{{\ttfamily arXiv:1408.4785 [hep-ph]}}.

\bibitem{Okada:2015vwh}
H.~Okada and Y.~Orikasa, ``{Radiative neutrino model with an inert triplet
  scalar},'' \href{http://dx.doi.org/10.1103/PhysRevD.94.055002}{{\em Phys.
  Rev.} {\bfseries D94} no.~5, (2016) 055002},
\href{http://arxiv.org/abs/1512.06687}{{\ttfamily arXiv:1512.06687 [hep-ph]}}.

\bibitem{Chowdhury:2015sla}
T.~A. Chowdhury and S.~Nasri, ``{Lepton Flavor Violation in the Inert Scalar
  Model with Higher Representations},''
  \href{http://dx.doi.org/10.1007/JHEP12(2015)040}{{\em JHEP} {\bfseries 12}
  (2015) 040},
\href{http://arxiv.org/abs/1506.00261}{{\ttfamily arXiv:1506.00261 [hep-ph]}}.

\bibitem{Diaz:2016udz}
M.~A. D\'iaz, N.~Rojas, S.~Urrutia-Quiroga, and J.~W.~F. Valle, ``{Heavy Higgs
  Boson Production at Colliders in the Singlet-Triplet Scotogenic Dark Matter
  Model},'' \href{http://dx.doi.org/10.1007/JHEP08(2017)017}{{\em JHEP}
  {\bfseries 08} (2017) 017},
\href{http://arxiv.org/abs/1612.06569}{{\ttfamily arXiv:1612.06569 [hep-ph]}}.

\bibitem{Ferreira:2016sbb}
P.~M. Ferreira, W.~Grimus, D.~Jurciukonis, and L.~Lavoura, ``{Scotogenic model
  for co-bimaximal mixing},''
  \href{http://dx.doi.org/10.1007/JHEP07(2016)010}{{\em JHEP} {\bfseries 07}
  (2016) 010},
\href{http://arxiv.org/abs/1604.07777}{{\ttfamily arXiv:1604.07777 [hep-ph]}}.

\bibitem{Ahriche:2016cio}
A.~Ahriche, K.~L. McDonald, and S.~Nasri, ``{The Scale-Invariant Scotogenic
  Model},'' \href{http://dx.doi.org/10.1007/JHEP06(2016)182}{{\em JHEP}
  {\bfseries 06} (2016) 182},
\href{http://arxiv.org/abs/1604.05569}{{\ttfamily arXiv:1604.05569 [hep-ph]}}.

\bibitem{vonderPahlen:2016cbw}
F.~von~der Pahlen, G.~Palacio, D.~Restrepo, and O.~Zapata, ``{Radiative Type
  III Seesaw Model and its collider phenomenology},''
  \href{http://dx.doi.org/10.1103/PhysRevD.94.033005}{{\em Phys. Rev.}
  {\bfseries D94} no.~3, (2016) 033005},
\href{http://arxiv.org/abs/1605.01129}{{\ttfamily arXiv:1605.01129 [hep-ph]}}.

\bibitem{Lu:2016dbc}
W.-B. Lu and P.-H. Gu, ``{Mixed Inert Scalar Triplet Dark Matter, Radiative
  Neutrino Masses and Leptogenesis},''
  \href{http://dx.doi.org/10.1016/j.nuclphysb.2017.09.005}{{\em Nucl. Phys.}
  {\bfseries B924} (2017) 279--311},
\href{http://arxiv.org/abs/1611.02106}{{\ttfamily arXiv:1611.02106 [hep-ph]}}.

\bibitem{Merle:2016scw}
A.~Merle, M.~Platscher, N.~Rojas, J.~W.~F. Valle, and A.~Vicente,
  ``{Consistency of WIMP Dark Matter as radiative neutrino mass messenger},''
  \href{http://dx.doi.org/10.1007/JHEP07(2016)013}{{\em JHEP} {\bfseries 07}
  (2016) 013},
\href{http://arxiv.org/abs/1603.05685}{{\ttfamily arXiv:1603.05685 [hep-ph]}}.

\bibitem{Rocha-Moran:2016enp}
P.~Rocha-Moran and A.~Vicente, ``{Lepton Flavor Violation in the
  singlet-triplet scotogenic model},''
  \href{http://dx.doi.org/10.1007/JHEP07(2016)078}{{\em JHEP} {\bfseries 07}
  (2016) 078},
\href{http://arxiv.org/abs/1605.01915}{{\ttfamily arXiv:1605.01915 [hep-ph]}}.

\bibitem{Chowdhury:2016mtl}
T.~A. Chowdhury and S.~Nasri, ``{The Sommerfeld Enhancement in the Scotogenic
  Model with Large Electroweak Scalar Multiplets},''
  \href{http://dx.doi.org/10.1088/1475-7516/2017/01/041}{{\em JCAP} {\bfseries
  1701} no.~01, (2017) 041},
\href{http://arxiv.org/abs/1611.06590}{{\ttfamily arXiv:1611.06590 [hep-ph]}}.

\bibitem{Fortes:2017ndr}
E.~C. F.~S. Fortes, A.~C.~B. Machado, J.~Monta\~{n}o, and V.~Pleitez, ``{Lepton
  masses and mixing in a scotogenic model},''
  \href{http://dx.doi.org/10.1016/j.physletb.2020.135289}{{\em Phys. Lett.}
  {\bfseries B803} (2020) 135289},
\href{http://arxiv.org/abs/1705.09414}{{\ttfamily arXiv:1705.09414 [hep-ph]}}.

\bibitem{Tang:2017rhv}
Y.-L. Tang, ``{Some Phenomenologies of a Simple Scotogenic Inverse Seesaw
  Model},'' \href{http://dx.doi.org/10.1103/PhysRevD.97.035020}{{\em Phys.
  Rev.} {\bfseries D97} no.~3, (2018) 035020},
\href{http://arxiv.org/abs/1709.07735}{{\ttfamily arXiv:1709.07735 [hep-ph]}}.

\bibitem{Guo:2018iix}
C.~Guo, S.-Y. Guo, and Y.~Liao, ``{Dark matter and LHC phenomenology of a scale
  invariant scotogenic model},''
  \href{http://dx.doi.org/10.1088/1674-1137/43/10/103102}{{\em Chin. Phys.}
  {\bfseries C43} no.~10, (2019) 103102},
\href{http://arxiv.org/abs/1811.01180}{{\ttfamily arXiv:1811.01180 [hep-ph]}}.

\bibitem{Rojas:2018wym}
N.~Rojas, R.~Srivastava, and J.~W.~F. Valle, ``{Simplest Scoto-Seesaw
  Mechanism},'' \href{http://dx.doi.org/10.1016/j.physletb.2018.12.014}{{\em
  Phys. Lett.} {\bfseries B789} (2019) 132--136},
\href{http://arxiv.org/abs/1807.11447}{{\ttfamily arXiv:1807.11447 [hep-ph]}}.

\bibitem{Aranda:2018lif}
A.~Aranda, C.~Bonilla, and E.~Peinado, ``{Dynamical generation of neutrino mass
  scales},'' \href{http://dx.doi.org/10.1016/j.physletb.2019.01.068}{{\em Phys.
  Lett.} {\bfseries B792} (2019) 40--42},
\href{http://arxiv.org/abs/1808.07727}{{\ttfamily arXiv:1808.07727 [hep-ph]}}.

\bibitem{Han:2019lux}
Z.-L. Han and W.~Wang, ``{Predictive Scotogenic Model with Flavor Dependent
  Symmetry},'' \href{http://dx.doi.org/10.1140/epjc/s10052-019-7033-8}{{\em
  Eur. Phys. J.} {\bfseries C79} no.~6, (2019) 522},
\href{http://arxiv.org/abs/1901.07798}{{\ttfamily arXiv:1901.07798 [hep-ph]}}.

\bibitem{Suematsu:2019kst}
D.~Suematsu, ``{Low scale leptogenesis in a hybrid model of the scotogenic type
  I and III seesaw models},''
  \href{http://dx.doi.org/10.1103/PhysRevD.100.055008}{{\em Phys. Rev.}
  {\bfseries D100} no.~5, (2019) 055008},
\href{http://arxiv.org/abs/1906.12008}{{\ttfamily arXiv:1906.12008 [hep-ph]}}.

\bibitem{Kang:2019sab}
S.~K. Kang, O.~Popov, R.~Srivastava, J.~W.~F. Valle, and C.~A. Vaquera-Araujo,
  ``{Scotogenic dark matter stability from gauged matter parity},''
  \href{http://dx.doi.org/10.1016/j.physletb.2019.135013}{{\em Phys. Lett.}
  {\bfseries B798} (2019) 135013},
\href{http://arxiv.org/abs/1902.05966}{{\ttfamily arXiv:1902.05966 [hep-ph]}}.

\bibitem{Pramanick:2019oxb}
S.~Pramanick, ``{Scotogenic S3 symmetric generation of realistic neutrino
  mixing},'' \href{http://dx.doi.org/10.1103/PhysRevD.100.035009}{{\em Phys.
  Rev.} {\bfseries D100} no.~3, (2019) 035009},
\href{http://arxiv.org/abs/1904.07558}{{\ttfamily arXiv:1904.07558 [hep-ph]}}.

\bibitem{Nomura:2019lnr}
T.~Nomura, H.~Okada, and O.~Popov, ``{A modular $A_4$ symmetric scotogenic
  model},'' \href{http://dx.doi.org/10.1016/j.physletb.2020.135294}{{\em Phys.
  Lett.} {\bfseries B803} (2020) 135294},
\href{http://arxiv.org/abs/1908.07457}{{\ttfamily arXiv:1908.07457 [hep-ph]}}.

\bibitem{Restrepo:2019ilz}
D.~Restrepo and A.~Rivera, ``{Phenomenological consistency of the
  singlet-triplet scotogenic model},''
\href{http://arxiv.org/abs/1907.11938}{{\ttfamily arXiv:1907.11938 [hep-ph]}}.

\bibitem{Rojas:2019llr}
N.~Rojas, R.~Srivastava, and J.~W.~F. Valle, ``{Scotogenic origin of the
  Inverse Seesaw Mechanism},''
\href{http://arxiv.org/abs/1907.07728}{{\ttfamily arXiv:1907.07728 [hep-ph]}}.

\bibitem{Avila:2019hhv}
I.~M. Ávila, V.~De~Romeri, L.~Duarte, and J.~W.~F. Valle, ``{Minimalistic
  scotogenic scalar dark matter},''
\href{http://arxiv.org/abs/1910.08422}{{\ttfamily arXiv:1910.08422 [hep-ph]}}.

\bibitem{Kumar:2019tat}
N.~Kumar, T.~Nomura, and H.~Okada, ``{Scotogenic neutrino mass with large
  $SU(2)_L$ multiplet fields},''
\href{http://arxiv.org/abs/1912.03990}{{\ttfamily arXiv:1912.03990 [hep-ph]}}.

\bibitem{Arbelaez:2020uer}
C.~Arbel\'aez, G.~Cottin, J.~C. Helo, and M.~Hirsch, ``{Long-lived charged
  particles and multi-lepton signatures from neutrino mass models},''
\href{http://arxiv.org/abs/2003.11494}{{\ttfamily arXiv:2003.11494 [hep-ph]}}.

\bibitem{Ma:2013yga}
E.~Ma, I.~Picek, and B.~Radov\v{c}i\'c, ``{New Scotogenic Model of Neutrino
  Mass with $U(1)_D$ Gauge Interaction},''
  \href{http://dx.doi.org/10.1016/j.physletb.2013.09.049}{{\em Phys. Lett.}
  {\bfseries B726} (2013) 744--746},
\href{http://arxiv.org/abs/1308.5313}{{\ttfamily arXiv:1308.5313 [hep-ph]}}.

\bibitem{Yu:2016lof}
J.-H. Yu, ``{Hidden Gauged U(1) Model: Unifying Scotogenic Neutrino and Flavor
  Dark Matter},'' \href{http://dx.doi.org/10.1103/PhysRevD.93.113007}{{\em
  Phys. Rev.} {\bfseries D93} no.~11, (2016) 113007},
\href{http://arxiv.org/abs/1601.02609}{{\ttfamily arXiv:1601.02609 [hep-ph]}}.

\bibitem{Kubo:2006rm}
J.~Kubo and D.~Suematsu, ``{Neutrino masses and CDM in a non-supersymmetric
  model},'' \href{http://dx.doi.org/10.1016/j.physletb.2006.11.005}{{\em Phys.
  Lett.} {\bfseries B643} (2006) 336--341},
\href{http://arxiv.org/abs/hep-ph/0610006}{{\ttfamily arXiv:hep-ph/0610006
  [hep-ph]}}.

\bibitem{Sierra:2014kua}
D.~Aristizabal~Sierra, M.~Dhen, C.~S. Fong, and A.~Vicente, ``{Dynamical flavor
  origin of $\mathbb{Z}_N$ symmetries},''
  \href{http://dx.doi.org/10.1103/PhysRevD.91.096004}{{\em Phys. Rev.}
  {\bfseries D91} no.~9, (2015) 096004},
\href{http://arxiv.org/abs/1412.5600}{{\ttfamily arXiv:1412.5600 [hep-ph]}}.

\bibitem{Hagedorn:2018spx}
C.~Hagedorn, J.~Herrero-Garc\'ia, E.~Molinaro, and M.~A. Schmidt,
  ``{Phenomenology of the Generalised Scotogenic Model with Fermionic Dark
  Matter},'' \href{http://dx.doi.org/10.1007/JHEP11(2018)103}{{\em JHEP}
  {\bfseries 11} (2018) 103},
\href{http://arxiv.org/abs/1804.04117}{{\ttfamily arXiv:1804.04117 [hep-ph]}}.

\bibitem{Bonilla:2019ipe}
C.~Bonilla, L.~M.~G. de~la Vega, J.~M. Lamprea, R.~A. Lineros, and E.~Peinado,
  ``{Fermion Dark Matter and Radiative Neutrino Masses from Spontaneous Lepton
  Number Breaking},'' \href{http://dx.doi.org/10.1088/1367-2630/ab7254}{{\em
  New J. Phys.} {\bfseries 22} no.~3, (2020) 033009},
\href{http://arxiv.org/abs/1908.04276}{{\ttfamily arXiv:1908.04276 [hep-ph]}}.

\bibitem{Ma:2017zyb}
E.~Ma, D.~Restrepo, and O.~Zapata, ``{Anomalous leptonic U(1) symmetry:
  Syndetic origin of the QCD axion, weak-scale dark matter, and radiative
  neutrino mass},'' \href{http://dx.doi.org/10.1142/S0217732318500244}{{\em
  Mod. Phys. Lett.} {\bfseries A33} (2018) 1850024},
\href{http://arxiv.org/abs/1706.08240}{{\ttfamily arXiv:1706.08240 [hep-ph]}}.

\bibitem{Carvajal:2018ohk}
C.~D.~R. Carvajal and O.~Zapata, ``{One-loop Dirac neutrino mass and mixed
  axion-WIMP dark matter},''
  \href{http://dx.doi.org/10.1103/PhysRevD.99.075009}{{\em Phys. Rev.}
  {\bfseries D99} no.~7, (2019) 075009},
\href{http://arxiv.org/abs/1812.06364}{{\ttfamily arXiv:1812.06364 [hep-ph]}}.

\bibitem{delaVega:2020jcp}
L.~M.~G. de~la Vega, N.~Nath, and E.~Peinado, ``{Dirac neutrinos from
  Peccei-Quinn symmetry: two examples},''
\href{http://arxiv.org/abs/2001.01846}{{\ttfamily arXiv:2001.01846 [hep-ph]}}.

\bibitem{Parida:2011wh}
M.~K. Parida, ``{Radiative Seesaw in SO(10) with Dark Matter},''
  \href{http://dx.doi.org/10.1016/j.physletb.2011.09.016}{{\em Phys. Lett.}
  {\bfseries B704} (2011) 206--210},
\href{http://arxiv.org/abs/1106.4137}{{\ttfamily arXiv:1106.4137 [hep-ph]}}.

\bibitem{Leite:2019grf}
J.~Leite, O.~Popov, R.~Srivastava, and J.~W.~F. Valle, ``{A theory for
  scotogenic dark matter stabilised by residual gauge symmetry},''
\href{http://arxiv.org/abs/1909.06386}{{\ttfamily arXiv:1909.06386 [hep-ph]}}.

\bibitem{Han:2019diw}
Z.-L. Han, R.~Ding, S.-J. Lin, and B.~Zhu, ``{Gauged $U(1)_{L_\mu -L_\tau }$
  scotogenic model in light of $R_{K^{(*)}}$ anomaly and AMS-02 positron
  excess},'' \href{http://dx.doi.org/10.1140/epjc/s10052-019-7526-5}{{\em Eur.
  Phys. J.} {\bfseries C79} no.~12, (2019) 1007},
\href{http://arxiv.org/abs/1908.07192}{{\ttfamily arXiv:1908.07192 [hep-ph]}}.

\bibitem{Wang:2019byi}
W.~Wang and Z.-L. Han, ``{The $U(1)_{B-3L_{\alpha}}$ Extended Scotogenic Models
  and One-texture-zeros of Neutrino Mass Matrices},''
  \href{http://arxiv.org/abs/1911.00819}{{\ttfamily arXiv:1911.00819
  [hep-ph]}}.

\bibitem{Hehn:2012kz}
D.~Hehn and A.~Ibarra, ``{A radiative model with a naturally mild neutrino mass
  hierarchy},'' \href{http://dx.doi.org/10.1016/j.physletb.2012.11.034}{{\em
  Phys. Lett.} {\bfseries B718} (2013) 988--991},
\href{http://arxiv.org/abs/1208.3162}{{\ttfamily arXiv:1208.3162 [hep-ph]}}.

\bibitem{Fuentes-Martin:2019dxt}
J.~Fuentes-Martín, M.~Reig, and A.~Vicente, ``{Strong $CP$ problem with
  low-energy emergent QCD: The 4321 case},''
  \href{http://dx.doi.org/10.1103/PhysRevD.100.115028}{{\em Phys. Rev.}
  {\bfseries D100} no.~11, (2019) 115028},
\href{http://arxiv.org/abs/1907.02550}{{\ttfamily arXiv:1907.02550 [hep-ph]}}.

\bibitem{tHooft:1979rat}
G.~'t~Hooft, ``{Naturalness, chiral symmetry, and spontaneous chiral symmetry
  breaking},'' \href{http://dx.doi.org/10.1007/978-1-4684-7571-5\_9}{{\em NATO
  Sci.\ Ser.\ B} {\bfseries 59} (1980) 135--157}.

\bibitem{Passarino:1978jh}
G.~Passarino and M.~J.~G. Veltman, ``{One Loop Corrections for $e^+ e^-$
  Annihilation Into $\mu^+ \mu^-$ in the Weinberg Model},''
\href{http://dx.doi.org/10.1016/0550-3213(79)90234-7}{{\em Nucl. Phys.}
  {\bfseries B160} (1979) 151--207}.

\bibitem{Merle:2015gea}
A.~Merle and M.~Platscher, ``{Parity Problem of the Scotogenic Neutrino
  Model},'' \href{http://dx.doi.org/10.1103/PhysRevD.92.095002}{{\em Phys.
  Rev.} {\bfseries D92} no.~9, (2015) 095002},
\href{http://arxiv.org/abs/1502.03098}{{\ttfamily arXiv:1502.03098 [hep-ph]}}.

\bibitem{Vicente:2015zba}
A.~Vicente, ``{Computer tools in particle physics},''
\href{http://arxiv.org/abs/1507.06349}{{\ttfamily arXiv:1507.06349 [hep-ph]}}.

\bibitem{Merle:2015ica}
A.~Merle and M.~Platscher, ``{Running of radiative neutrino masses: the
  scotogenic model - revisited},''
  \href{http://dx.doi.org/10.1007/JHEP11(2015)148}{{\em JHEP} {\bfseries 11}
  (2015) 148},
\href{http://arxiv.org/abs/1507.06314}{{\ttfamily arXiv:1507.06314 [hep-ph]}}.

\bibitem{Lindner:2016kqk}
M.~Lindner, M.~Platscher, C.~E. Yaguna, and A.~Merle, ``{Fermionic WIMPs and
  vacuum stability in the scotogenic model},''
  \href{http://dx.doi.org/10.1103/PhysRevD.94.115027}{{\em Phys. Rev.}
  {\bfseries D94} no.~11, (2016) 115027},
\href{http://arxiv.org/abs/1608.00577}{{\ttfamily arXiv:1608.00577 [hep-ph]}}.

\bibitem{Casas:2001sr}
J.~Casas and A.~Ibarra, ``{Oscillating neutrinos and $\mu \to e, \gamma$},''
  \href{http://dx.doi.org/10.1016/S0550-3213(01)00475-8}{{\em Nucl.\ Phys.\ B}
  {\bfseries 618} (2001) 171--204},
  \href{http://arxiv.org/abs/hep-ph/0103065}{{\ttfamily arXiv:hep-ph/0103065}}.

\bibitem{Toma:2013zsa}
T.~Toma and A.~Vicente, ``{Lepton Flavor Violation in the Scotogenic Model},''
  \href{http://dx.doi.org/10.1007/JHEP01(2014)160}{{\em JHEP} {\bfseries 01}
  (2014) 160}, \href{http://arxiv.org/abs/1312.2840}{{\ttfamily arXiv:1312.2840
  [hep-ph]}}.

\bibitem{Vicente:2014wga}
A.~Vicente and C.~E. Yaguna, ``{Probing the scotogenic model with lepton flavor
  violating processes},'' \href{http://dx.doi.org/10.1007/JHEP02(2015)144}{{\em
  JHEP} {\bfseries 02} (2015) 144},
  \href{http://arxiv.org/abs/1412.2545}{{\ttfamily arXiv:1412.2545 [hep-ph]}}.

\bibitem{Cordero-Carrion:2018xre}
I.~Cordero-Carrión, M.~Hirsch, and A.~Vicente, ``{Master Majorana neutrino
  mass parametrization},''
  \href{http://dx.doi.org/10.1103/PhysRevD.99.075019}{{\em Phys.\ Rev.\ D}
  {\bfseries 99} no.~7, (2019) 075019},
  \href{http://arxiv.org/abs/1812.03896}{{\ttfamily arXiv:1812.03896
  [hep-ph]}}.

\bibitem{Cordero-Carrion:2019qtu}
I.~Cordero-Carrión, M.~Hirsch, and A.~Vicente, ``{General parametrization of
  Majorana neutrino mass models},''
  \href{http://arxiv.org/abs/1912.08858}{{\ttfamily arXiv:1912.08858
  [hep-ph]}}.

\bibitem{deSalas:2017kay}
P.~de~Salas, D.~Forero, C.~Ternes, M.~Tortola, and J.~Valle, ``{Status of
  neutrino oscillations 2018: 3$\sigma$ hint for normal mass ordering and
  improved CP sensitivity},''
  \href{http://dx.doi.org/10.1016/j.physletb.2018.06.019}{{\em Phys.\ Lett.\ B}
  {\bfseries 782} (2018) 633--640},
  \href{http://arxiv.org/abs/1708.01186}{{\ttfamily arXiv:1708.01186
  [hep-ph]}}.

\bibitem{Quiros:1999jp}
M.~Quiros, ``{Finite temperature field theory and phase transitions},'' in {\em
  {Proceedings, Summer School in High-energy physics and cosmology: Trieste,
  Italy, June 29-July 17, 1998}}, pp.~187--259.
\newblock 1999.
\newblock
\href{http://arxiv.org/abs/hep-ph/9901312}{{\ttfamily arXiv:hep-ph/9901312
  [hep-ph]}}.
\newblock

\bibitem{Gil:2012ya}
G.~Gil, P.~Chankowski, and M.~Krawczyk, ``{Inert Dark Matter and Strong
  Electroweak Phase Transition},''
  \href{http://dx.doi.org/10.1016/j.physletb.2012.09.052}{{\em Phys. Lett.}
  {\bfseries B717} (2012) 396--402},
\href{http://arxiv.org/abs/1207.0084}{{\ttfamily arXiv:1207.0084 [hep-ph]}}.

\bibitem{Blinov:2015vma}
N.~Blinov, S.~Profumo, and T.~Stefaniak, ``{The Electroweak Phase Transition in
  the Inert Doublet Model},''
  \href{http://dx.doi.org/10.1088/1475-7516/2015/07/028}{{\em JCAP} {\bfseries
  1507} no.~07, (2015) 028},
\href{http://arxiv.org/abs/1504.05949}{{\ttfamily arXiv:1504.05949 [hep-ph]}}.

\bibitem{Linde:2005ht}
A.~D. Linde, ``{Particle physics and inflationary cosmology},'' {\em Contemp.
  Concepts Phys.} {\bfseries 5} (1990) 1--362,
\href{http://arxiv.org/abs/hep-th/0503203}{{\ttfamily arXiv:hep-th/0503203
  [hep-th]}}.

\bibitem{Staub:2008uz}
F.~Staub, ``{SARAH},''
\href{http://arxiv.org/abs/0806.0538}{{\ttfamily arXiv:0806.0538 [hep-ph]}}.

\bibitem{Staub:2009bi}
F.~Staub, ``{From Superpotential to Model Files for FeynArts and
  CalcHep/CompHep},'' \href{http://dx.doi.org/10.1016/j.cpc.2010.01.011}{{\em
  Comput. Phys. Commun.} {\bfseries 181} (2010) 1077--1086},
\href{http://arxiv.org/abs/0909.2863}{{\ttfamily arXiv:0909.2863 [hep-ph]}}.

\bibitem{Staub:2010jh}
F.~Staub, ``{Automatic Calculation of supersymmetric Renormalization Group
  Equations and Self Energies},''
  \href{http://dx.doi.org/10.1016/j.cpc.2010.11.030}{{\em Comput. Phys.
  Commun.} {\bfseries 182} (2011) 808--833},
\href{http://arxiv.org/abs/1002.0840}{{\ttfamily arXiv:1002.0840 [hep-ph]}}.

\bibitem{Staub:2012pb}
F.~Staub, ``{SARAH 3.2: Dirac Gauginos, UFO output, and more},''
  \href{http://dx.doi.org/10.1016/j.cpc.2013.02.019}{{\em Comput. Phys.
  Commun.} {\bfseries 184} (2013) 1792--1809},
\href{http://arxiv.org/abs/1207.0906}{{\ttfamily arXiv:1207.0906 [hep-ph]}}.

\bibitem{Staub:2013tta}
F.~Staub, ``{SARAH 4 : A tool for (not only SUSY) model builders},''
  \href{http://dx.doi.org/10.1016/j.cpc.2014.02.018}{{\em Comput. Phys.
  Commun.} {\bfseries 185} (2014) 1773--1790},
\href{http://arxiv.org/abs/1309.7223}{{\ttfamily arXiv:1309.7223 [hep-ph]}}.

\bibitem{Kannike:2016fmd}
K.~Kannike, ``{Vacuum Stability of a General Scalar Potential of a Few
  Fields},'' \href{http://dx.doi.org/10.1140/epjc/s10052-016-4160-3}{{\em Eur.
  Phys. J. C} {\bfseries 76} no.~6, (2016) 324},
  \href{http://arxiv.org/abs/1603.02680}{{\ttfamily arXiv:1603.02680
  [hep-ph]}}. [Erratum: Eur.Phys.J.C 78, 355 (2018)].

\bibitem{Ivanov:2018jmz}
I.~P. Ivanov, M.~Köpke, and M.~Mühlleitner, ``{Algorithmic
  Boundedness-From-Below Conditions for Generic Scalar Potentials},''
  \href{http://dx.doi.org/10.1140/epjc/s10052-018-5893-y}{{\em Eur. Phys. J.}
  {\bfseries C78} no.~5, (2018) 413},
\href{http://arxiv.org/abs/1802.07976}{{\ttfamily arXiv:1802.07976 [hep-ph]}}.

\bibitem{Kannike:2012pe}
K.~Kannike, ``{Vacuum Stability Conditions From Copositivity Criteria},''
  \href{http://dx.doi.org/10.1140/epjc/s10052-012-2093-z}{{\em Eur. Phys. J.}
  {\bfseries C72} (2012) 2093},
\href{http://arxiv.org/abs/1205.3781}{{\ttfamily arXiv:1205.3781 [hep-ph]}}.

\bibitem{KAPLAN2000203}
W.~Kaplan, ``A test for copositive matrices,''
  \href{http://dx.doi.org/10.1016/S0024-3795(00)00138-5}{{\em Linear Algebra
  and its Applications} {\bfseries 313} no.~1, (2000) 203 -- 206}.

\bibitem{Yang-Li}
S.-j. Yang and X.-x. Li, ``Algorithms for determining the copositivity of a
  given symmetric matrix,''
  \href{http://dx.doi.org/10.1016/j.laa.2008.07.028}{{\em Linear Algebra and
  its Applications} {\bfseries 430} no.~2, (2009) 609 -- 618}.

\bibitem{Yang2010}
S.-J. Yang, C.-Q. Xu, and X.-X. Li, ``A note on algorithms for determining the
  copositivity of a given symmetric matrix,''
  \href{http://dx.doi.org/10.1155/2010/498631}{{\em Journal of Inequalities and
  Applications} {\bfseries 2010} (2009) 498631}.

\bibitem{Faro:2019vcd}
F.~S. Faro and I.~P. Ivanov, ``{Boundedness from below in the $U(1)\times U(1)$
  three-Higgs-doublet model},''
  \href{http://dx.doi.org/10.1103/PhysRevD.100.035038}{{\em Phys. Rev.}
  {\bfseries D100} no.~3, (2019) 035038},
\href{http://arxiv.org/abs/1907.01963}{{\ttfamily arXiv:1907.01963 [hep-ph]}}.

\end{thebibliography}\endgroup

\end{document}